\documentclass{article}
\usepackage[a4paper]{geometry}
\usepackage{graphicx} 
\usepackage{amsmath}
\usepackage{stackengine}
\usepackage[dvipsnames]{xcolor}
\usepackage{amsfonts}
\usepackage{subcaption}
\usepackage{natbib}
\usepackage{hyperref}
\usepackage{doi}
\usepackage{algorithm}
\usepackage{algpseudocode}
\setcitestyle{authoryear, open={(},close={)}}
\usepackage{comment}
\usepackage{datetime}
\usepackage{float}
\usepackage{mathabx}
\usepackage{soul}
\usepackage{tikz}
\usepackage{a4wide}
\usetikzlibrary{arrows.meta, positioning,decorations.pathreplacing}

\newtheorem{remark}{Remark}

\title{Splitting-based extended Kalman filtering for parameter inference in the stochastic Jansen--Rit neural mass model}

\author{Simon Pauli\footnotemark[1]\thanks{Institute of Applied Statistics, Johannes Kepler University Linz (simon.pauli@jku.at)}, Irene Tubikanec\footnotemark[2]\thanks{Institute of Applied Statistics, Johannes Kepler University Linz (irene.tubikanec@jku.at)}, Andreas Futschik\footnotemark[3]\thanks{Institute of Applied Statistics, Johannes Kepler University Linz (andreas.futschik@jku.at)}}

\date{}

\newdateformat{monthyeardate}{
  \monthname[\THEMONTH], \THEYEAR}

\newcommand{\sigm}{\text{Sigm}}

\begin{document}

\maketitle

\begin{abstract}

In this article, we develop a computationally efficient approach to parameter inference for the stochastic Jansen--Rit neural mass model (JR-NMM), combining the extended Kalman filter (EKF) with structure-preserving numerical splitting methods. The stochastic JR-NMM is a six-dimensional stochastic differential equation (SDE) that describes the activity of a population of neurons in the brain and can produce EEG-like signals. Parameter inference for this model is challenging because of its hypoellipticity, nonlinear dynamics, lack of closed-form transition distributions, and partial observation of the model state. Our approach uses the EKF to construct an approximate likelihood for parameter estimation, based on the Ornstein--Uhlenbeck (OU) splitting method, which combines the linear drift dynamics and diffusion term into an exactly solvable stochastic subequation. The resulting OU-EKF benefits from a full-rank transition covariance matrix and accounts for a nontrivial process-noise Jacobian in the covariance prediction.

In a simulation study, the OU-EKF outperforms both an alternative splitting-based EKF and an EKF based on the standard Euler--Maruyama method. Further experiments demonstrate its robustness across different observation settings. Compared with a state-of-the-art approximate Bayesian computation approach, the OU-EKF achieves higher parameter estimation accuracy at substantially lower computational cost. Applied to real EEG data exhibiting alpha-rhythmic activity, the method yields physiologically plausible parameter estimates that are robust across different treatments of model discrepancy and consistent with findings in the literature. The proposed methodology may also hold promise beyond the stochastic JR-NMM, particularly for other partially observed SDEs for which suitable numerical splitting methods can be constructed.
\end{abstract}

\subsubsection*{Keywords} Parameter inference, Extended Kalman filter, Numerical splitting methods, Jansen--Rit neural mass model, EEG data.

\subsubsection*{MSC Subject classifications} 62M20, 65C30, 60G35, 60H10     

\section{Introduction}

Parameter inference for multidimensional stochastic differential equations (SDEs) is challenging, particularly when the dynamics are nonlinear, the transition distributions are not available in closed form, the diffusion is degenerate, and the system is partially observed. The latter makes inference particularly difficult when only a low-dimensional function of the state variables is observed. These difficulties are especially relevant in biomedical applications, where complex, multidimensional mechanistic models are used to describe physiological processes.

Neural mass models constitute an important class of such mathematical models. They describe the activity of large neuronal populations and have been studied since the 1970s (see, e.g., \cite{WILSON19721,LopesDaSilva1974,LOPESDASILVA1976,Freeman1978}). These models have been used, for example, in neuroimaging \citep{friston2005models}, in the investigation of electroencephalogram (EEG) rhythms \citep{DAVID2003,Sotero2007}, and for the understanding of epileptic brain dynamics \citep{wendling2000relevance,WENDLING2016}. For further insights into their development and possible applications, see \mbox{\cite{Coombes2019}}. 

In this work, we focus on the Jansen--Rit neural mass model (JR-NMM), which was introduced to describe human EEG rhythms and visual evoked potentials \citep{jansenRit}. The JR-NMM describes the average activity of a population of neurons, such as a cortical column, through three interacting subpopulations (pyramidal neurons, excitatory interneurons, and inhibitory interneurons) and can reproduce EEG-like activity, including background rhythms, alpha oscillations, and epileptic activity. The original JR-NMM is formulated as a system of three coupled nonlinear second-order ordinary differential equations (ODEs), describing the dynamics of the three neuronal subpopulations, or equivalently as a six-dimensional first-order system. It includes an external input representing activity from surrounding or more distant neuronal populations, which is commonly modeled as a noisy input function. To provide a mathematically well-defined description of this stochastic forcing, \cite{haraldPaper} reformulated the JR-NMM as a six-dimensional SDE. This rigorous framework enabled the analysis of the model's structural properties (e.g., moment bounds, hypoellipticity and geometric ergodicity) using established tools from stochastic analysis, as well as the development and analysis of structure-preserving numerical methods for its approximation.

Beyond the theoretical and numerical analysis of the stochastic JR-NMM, an important question is how to infer its parameters from observed data. So far, parameter inference for the stochastic JR-NMM has been investigated using likelihood-free methods based on approximate Bayesian computation (ABC) \citep{ireneSpringer,ireneNetwork}. However, ABC methods are computationally expensive because they require many repeated simulations of the underlying stochastic model. Filtering methods provide a natural alternative for parameter inference in partially observed SDEs. They use the available observations sequentially to estimate the unobserved state and quantify uncertainty about its evolution. For nonlinear models, the extended Kalman filter (EKF) offers a computationally efficient approach based on Gaussian approximations and local linearization of the dynamics. These approximations also provide a likelihood-based criterion for estimating unknown model parameters. Such EKF-based parameter estimation approaches have been investigated for discretely observed SDEs (see,~e.g., \cite{singer2002parameter,kristensen2004greybox,GU2014,kleppe2014,sarkkaBook}).

A key difficulty in applying the EKF arises when the SDE cannot be solved exactly, so its transition distributions are not available in closed form. In this case, the EKF state prediction requires a numerical approximation of the model dynamics between consecutive observation times. The choice of numerical method therefore plays a crucial role in the filtering framework, affecting both the predicted states and their covariance matrices and, consequently, the approximate likelihood used for parameter estimation. A standard choice is the Euler--Maruyama~(EM) method \citep{KloedenPlaten}, an extension of Euler's method for ODEs to SDEs. Its simplicity makes it a natural starting point for constructing a filter. However, it is well known from the literature that this method may not preserve important structural properties of SDEs (see, e.g., \cite{Mattingly2002,Ditlevsen2019,Buckwar2022}), especially for oscillatory systems \citep{Strommen2004,Cohen2012}, and can thus require prohibitively small time steps to approximate the system dynamics accurately. This is particularly relevant for the stochastic JR-NMM, whose dynamics combine nonlinear interactions with oscillatory behavior \citep{haraldPaper}.

Numerical splitting methods, specifically constructed to preserve important qualitative properties of the underlying model, offer an alternative to standard EM discretization. Originally developed for ODEs (see, e.g., \cite{Mclachlan2002,Blanes2009}), splitting methods have since been substantially extended to SDEs (see, e.g., \cite{Misawa2001,Ableidinger2016,Alamo2016,Brehier2018,Chevallier2020,KellyLord2023,Fosteretal2022,JansLord2025}). The basic idea of these methods is to decompose an SDE that cannot be solved exactly into simpler subequations whose solutions are known explicitly. By composing the corresponding solution operators in a suitable order at each time step, one obtains a numerical approximation of the original SDE. For the stochastic JR-NMM, \citet{haraldPaper} developed two such splitting methods: a Wiener (WIE) and an Ornstein--Uhlenbeck (OU) splitting method. The WIE method treats the diffusion term separately from the linear drift dynamics. In contrast, the OU method combines the linear drift dynamics and the diffusion term into an exactly solvable stochastic subproblem. In \citet{haraldPaper}, both methods preserve important structural properties of the model, including moment bounds and geometric ergodicity, in contrast to the standard EM method. Although the two splitting methods exhibit similar properties from the perspective of stochastic numerical analysis, we show that their different treatment of the noise term has important implications for their use within an EKF.

In this paper, we combine the EKF approach with structure-preserving numerical splitting methods for parameter inference in the stochastic JR-NMM. Specifically, we construct and compare three EKF-based parameter estimation methods, using the EM, WIE, and OU discretizations for their respective transition approximations. We refer to these methods as \textbf{EM-EKF}, \textbf{WIE-EKF}, and \textbf{OU-EKF}, respectively. A central methodological contribution is deriving the state and process-noise Jacobians needed to incorporate the splitting transitions into the EKF prediction equations. In particular, our analysis reveals two important advantages of the OU formulation. First, the OU method yields a full-rank transition covariance matrix. Second, its treatment of process noise yields a nontrivial noise Jacobian, which we explicitly incorporate into the propagation of predictive uncertainty. These features distinguish the OU-EKF from the corresponding EM-EKF and WIE-EKF formulations and motivate an investigation into how numerical discretization affects the performance of EKF-based parameter inference.

We assess the performance of the three methods in a simulation study, comparing their ability to recover relevant parameters of the stochastic JR-NMM. The results show that both splitting-based methods outperform the EM-EKF, with the OU-EKF achieving the best inferential results. We then examine the OU-EKF in more detail, assessing its robustness to observation step size and time horizon and comparing it with the state-of-the-art ABC approach. 

Finally, we apply the OU-EKF to real EEG data exhibiting alpha-rhythmic activity to estimate the parameters of the stochastic JR-NMM. In particular, we investigate how different treatments of model discrepancy affect the resulting parameter estimates and the agreement between the observed signal and the fitted model. Overall, the application demonstrates that the OU-EKF can successfully estimate parameters of the stochastic JR-NMM from real EEG data, yielding plausible results consistent with established findings on alpha rhythms in the literature.

\paragraph{Organization of the paper.} The remainder of the paper is organized as follows. Section~\ref{sec:ekf} reviews the EKF approach and the associated approximate likelihood criterion. Section~\ref{sec:jr} introduces the stochastic JR-NMM and its numerical discretization methods. In Section~\ref{sec:jrEkf}, we derive the EKF formulations for the three discretization methods, with particular emphasis on the OU-EKF and its Jacobians. Section~\ref{sec:simulation} presents the simulation study, including a comparison of the three methods, an investigation of the OU-EKF's robustness, and a comparison with the state-of-the-art ABC approach.  Section~\ref{sec:application} applies the OU-EKF to real EEG data and examines different approaches to accounting for model discrepancy. Finally, Section~\ref{sec:discussion} discusses the main findings and possible extensions.

\paragraph{Further related literature.} Parameter inference for the original JR-NMM and related neural mass models has been investigated using several approaches. Closely related to our work, \citet{liuGao} employed the unscented Kalman filter (UKF), while \citet{Yokoyama2025Nov} used an ensemble Kalman filtering approach to investigate cortical excitation-inhibition balance. Other approaches include deep learning \citep{tilwani2025deepjansenritparameterinference} and genetic algorithms based on spectral characteristics \citep{app11020677}. None of these studies considers the SDE formulation of the JR-NMM investigated here or uses numerical splitting methods for parameter~inference. 

For the stochastic JR-NMM, the ABC-based parameter estimation approaches mentioned above employ the OU splitting method to generate simulated trajectories from the model. These approaches differ fundamentally from our EKF-based method: ABC compares simulated and observed data without evaluating a likelihood, whereas our method constructs an approximate likelihood. 

More generally, the use of numerical splitting methods for parameter inference in SDEs has only recently received attention. In particular, \citet{pilipovic2024,pilipovic2025} developed likelihood-based parameter estimation methods using splitting schemes, considering fully observed SDEs in the former and partially observed second-order SDEs in the latter. However, neither approach directly applies to the stochastic JR-NMM, where observations consist only of a one-dimensional function of two state variables. Splitting methods have also been employed in simulation-based inference approaches, such as ABC \citep{Samson2025,tamborrino2025arxiv}, in addition to the previously mentioned studies on the stochastic JR-NMM. Furthermore, splitting-based filtering for the stochastic JR-NMM has been investigated using the WIE method \citep{haraldDiss}. However, this work focused on state reconstruction under the assumption of known model parameters and did not address parameter estimation.

Our work thus introduces a novel combination of numerical splitting methods and EKF-based approximate likelihood inference for the stochastic JR-NMM, with particular emphasis on the OU splitting method. The proposed methodology may also hold promise beyond the stochastic JR-NMM, particularly for other SDE models for which suitable splitting methods can be constructed.

\paragraph{Notation.} We denote by $0_d$ the $d$-dimensional vector of zeros, and by
$\mathbb{O}_d$ the $d\times d$ matrix of zeros. $\mathbb{I}_d$ is the
$d\times d$ identity matrix, and $\operatorname{diag}(a_1,\ldots,a_d)$ a $d\times d$ diagonal matrix with diagonal entries $a_1,\ldots,a_d$. Moreover, $^\top$ denotes the transpose and $\| \cdot \|$ the Euclidean norm.  
\section{Extended Kalman filter-based parameter inference}
\label{sec:ekf}

The extended Kalman filter (EKF) is a widely used tool for nonlinear filtering tasks that relies on local linearization of the system dynamics around the current estimate of the state mean. In Section \ref{sec:ekf:1}, we briefly recall the general EKF framework before describing how to use it to construct an approximate likelihood function for parameter inference in Section \ref{sec:innovationLikelihood}. Finally, Section \ref{sec:uncertainty} introduces a measure for quantifying uncertainty in the resulting parameter estimates. This section largely follows \cite{sarkkaBook}.

\subsection{Extended Kalman filter}
\label{sec:ekf:1}

Consider a partially observed stochastic dynamical system with initial state $X_0 \in\mathbb{R}^{p}$, having mean $m_0$ and covariance matrix $P_0$. On a discrete observation grid, the state transitions and observations are given by
\begin{equation}
\label{eq:discreteStateSpace}
    X_{n} = f(X_{n-1},\xi_{n}), 
    \qquad
    Y_{n} = h(X_{n},r_{n}), \qquad n=1,\ldots,N,
\end{equation}
where $\{X_n\}_{n=0}^N$, with $X_n \in\mathbb{R}^{p}$, denotes the latent state sequence, and $\{Y_n\}_{n=1}^N$, with $Y_n\in\mathbb{R}^{d}$, 
the observation sequence. The random variables $\xi_n\in\mathbb R^p$ and $r_n\in\mathbb R^d$ represent process and observation noise,
respectively. Specifically, we assume
\[
    \xi_n\sim\mathcal{N}(0_p,\mathcal C_n),
    \qquad
    r_n\sim\mathcal{N}(0_d,R_n),
\]
where all noise variables are mutually independent and independent of the initial state $X_0$.

For each $n=1,\ldots,N$, let $y_{1:n}=(y_1,\ldots,y_n)$ denote the given observations up to time $n$, i.e., the observed realizations of the random variables $Y_1,\ldots,Y_n$. The EKF approximates the conditional distribution of the latent state $X_n$, given these observations, by a Gaussian distribution of the form
\[
    X_n\mid y_{1:n}
    \approx
    \mathcal{N}(m_n,P_n), 
\]
where $m_n$ and $P_n$ denote the corresponding approximate conditional mean and covariance matrix, respectively.

To compute these approximate conditional means and covariance matrices, the EKF starts from the initial mean $m_0$ and covariance matrix $P_0$ and proceeds recursively for $n=1,\ldots,N$ as follows. First, the predicted mean is obtained by propagating the previous filtered mean through~$f$, evaluated at $\xi_n=0_p$:
\begin{equation}
\label{eq:ekfPredictionMean}
    \bar m_n
    =
    f(m_{n-1},0_p).
\end{equation}
This transition is then linearized around $(m_{n-1},0_p)$. Define the Jacobians
\begin{equation}
\label{eq:ekfJacobians}
    F_{x,n}
    =
    \left.
    \frac{\partial f(x,\xi)}
         {\partial x}
    \right|_{x=m_{n-1},\,\xi=0_p},
    \qquad
    F_{\xi,n}
    =
    \left.
    \frac{\partial f(x,\xi)}
         {\partial \xi}
    \right|_{x=m_{n-1},\,\xi=0_p}.
\end{equation}
The predicted covariance is given by
\begin{equation}
\label{eq:ekfPredictionCov}
    \bar P_n
    =
    F_{x,n}P_{n-1}F_{x,n}^{\top}
    +
    F_{\xi,n}\mathcal C_nF_{\xi,n}^{\top},
\end{equation}
and the predicted mean is passed through $h$, yielding the predicted observation
\begin{equation}
\label{eq:ekfObservationPrediction}
    \bar y_n
    =
    h(\bar m_n,0_d).
\end{equation}
Defining
\begin{equation}
\label{eq:ekfObservationJacobians}
    H_{x,n}
    =
    \left.
    \frac{\partial h(x,r)}
         {\partial x}
    \right|_{x=\bar m_n,\,r=0_d},
    \qquad
    H_{r,n}
    =
    \left.
    \frac{\partial h(x,r)}
         {\partial r}
    \right|_{x=\bar m_n,\,r=0_d},
\end{equation}
the \textit{innovation} (one-step prediction error) and its covariance matrix are given by
\begin{align}
\label{eq:innovation}
    \nu_n
    &=
    y_n-\bar y_n,\\
\label{eq:innovationCov}
    S_n
    &=
    H_{x,n}\bar P_nH_{x,n}^{\top}
    +
    H_{r,n}R_nH_{r,n}^{\top}.
\end{align}
Moreover, the so-called Kalman gain is given by
\begin{equation}
\label{eq:kalmanGain}
    K_n
    =
    \bar P_nH_{x,n}^{\top}S_n^{-1}.
\end{equation}
Finally, the filtered mean update becomes
\begin{equation}
\label{eq:ekfUpdateMean}
    m_n
    =
    \bar m_n+K_n\nu_n,
\end{equation}
and the covariance update is given by 
\begin{equation}
\label{eq:josephForm}
    P_n
    =
    (\mathbb I_p-K_nH_{x,n})\bar P_n(\mathbb I_p-K_nH_{x,n})^{\top}
    +
    K_nH_{r,n}R_n H_{r,n}^{\top}K_n^{\top}.
\end{equation}
For the covariance update, we use the Joseph form \citep{josephFormBook}, which is algebraically equivalent to the standard simplified covariance update \citep{sarkkaBook}
in exact arithmetic but offers improved numerical stability by better preserving symmetry and positive semidefiniteness. In our
implementation, the simplified covariance update occasionally resulted in
invalid covariance matrices, motivating the use of \eqref{eq:josephForm}.

\subsection{Approximate innovation likelihood}
\label{sec:innovationLikelihood}

We now consider estimating an unknown parameter vector $\theta$, which governs the dynamics of the stochastic dynamical system, from the observations $y_{1:N}=\{y_1,\ldots,y_N\}$. 

The parameter vector $\theta$ may enter the state transition function $f$, the observation function~$h$, and the covariance matrices $\mathcal C_n$ and $R_n$ of the process and observation noise, respectively. Consequently, the quantities in the EKF recursion presented in the previous subsection, \eqref{eq:ekfPredictionMean}--\eqref{eq:josephForm}, may also depend on $\theta$. For each fixed $\theta$, the filter thus generates a corresponding sequence of innovations $\{\nu_n(\theta)\}_{n=1}^N$ and innovation covariance matrices $\{S_n(\theta)\}_{n=1}^N$.

Due to the generally nonlinear state transition and observation functions, the conditional distributions of the latent states arising in the filtering problem need not be Gaussian. The EKF approximates these distributions with Gaussian distributions by locally linearizing the corresponding functions. Consequently, given the observations up to time $n-1$, the innovation at time $n$ is approximately Gaussian with zero mean and covariance matrix $S_n(\theta)$:
\begin{equation}
\label{eq:innovationApprox}
    \nu_n(\theta)
    \approx
    \mathcal{N}\!\left(0_d,S_n(\theta)\right).
\end{equation}

This motivates the construction of an innovation-based approximate log-likelihood 
by summing the Gaussian log-densities of the
one-step prediction errors:
\begin{equation}
\label{eq:quasiLikelihood}
    \ell(\theta\mid y_{1:N})
    =
    -\frac{1}{2}
    \sum_{n=1}^{N}
    \left[
        d\log(2\pi)
        +\log\left|S_n(\theta)\right|
        +\nu_n(\theta)^{\top}
         S_n(\theta)^{-1}
         \nu_n(\theta)
    \right].
\end{equation}
Note that, due to the recursive nature of the EKF, the $n$th summand in \eqref{eq:quasiLikelihood} depends on the observations $y_{1:n}$, with the current observation $y_n$ entering through the innovation $\nu_n(\theta)$ (see~\eqref{eq:innovation}). We refer to \eqref{eq:quasiLikelihood} as an \textit{approximate} log-likelihood because it is based on the EKF approximation rather than the exact conditional distributions. In general, it therefore differs from the exact log-likelihood of the stochastic dynamical system.

The estimator for the unknown parameter vector $\theta$ is then defined as the maximizer of the approximate innovation log-likelihood:
\begin{equation}
\label{eq:qMLE}
    \hat\theta
    =
    \operatorname*{arg\,max}_{\theta}
    \ell(\theta\mid y_{1:N}).
\end{equation}
Since the innovations and their covariance matrices depend on $\theta$ through the recursive EKF equations, the maximizer generally cannot be obtained in closed form. We therefore compute $\hat\theta$ by numerical optimization. Specifically, we use the R function \texttt{optim} \citep{Rlang} with the L-BFGS-B method, which is suited for bounded optimization problems. This is particularly relevant since most parameters considered are non-negative, having a natural lower bound at zero. The likelihood evaluation is deterministic and implemented in \texttt{C++} via \texttt{Rcpp} \citep{RcppPack}, allowing us to combine the computational efficiency of \texttt{C++} with the optimization routines available in \texttt{R}. Further implementation details are provided in the accompanying code (see the Software statement at the end of the manuscript).

\subsection{Uncertainty quantification}
\label{sec:uncertainty}

In addition to point estimation via \eqref{eq:qMLE}, we aim to quantify
uncertainty in the estimated parameters. Since the approximate innovation
likelihood is based on the EKF approximation, the usual information-matrix
equality need not hold. Consequently, the inverse Hessian alone may not
provide an adequate estimate of the sampling covariance of the parameter
estimator. We therefore employ a sandwich covariance estimator, which
accounts for differences between the curvature of the objective function
and the variability of its score contributions
\citep{white1982}.

To account for the positivity constraints on the parameters, we work on
the logarithmic scale. Define
\[
    \phi = \log(\theta),
    \qquad
    \hat\phi = \log(\hat\theta),
\]
where the logarithm is applied componentwise and $\hat\theta$ is the
parameter estimate obtained from \eqref{eq:qMLE}. We write the negative
approximate log-likelihood as a sum of its one-step contributions,
\[
    \mathcal L(\phi)
    =
    -\ell(\exp(\phi)\mid y_{1:N})
    =
    \sum_{n=1}^{N} q_n(\phi),
\]
where
\[
    q_n(\phi)
    =
    \frac{1}{2}
    \left[
        d\log(2\pi)
        +\log|S_n(\exp(\phi))|
        +\nu_n(\exp(\phi))^\top
         S_n(\exp(\phi))^{-1}
         \nu_n(\exp(\phi))
    \right]
\]
denotes the negative log-likelihood contribution associated with the
$n$th innovation.

The first component of the sandwich estimator is the Hessian of the
negative approximate log-likelihood,
\[
    \boldsymbol{H}
    =
    \left.
    \nabla_\phi^2 \mathcal L(\phi)
    \right|_{\phi=\hat\phi}.
\]
The second component is based on the covariance of the individual
score contributions
\[
    g_n(\phi)
    =
    \nabla_\phi q_n(\phi),
    \qquad n=1,\ldots,N.
\]
Both the Hessian and the individual score contributions are evaluated
numerically using central finite differences, with step size
$h=10^{-3}$.

Since the individual score contributions may exhibit temporal
dependence, we estimate their long-run covariance using a
heteroskedasticity and autocorrelation consistent (HAC) estimator
following \cite{neweyWest1987}. In particular, we
employ the Bartlett kernel, which assigns decreasing weights to
successive lagged covariance contributions.

Let
\[
    \bar g
    =
    \frac{1}{N}\sum_{n=1}^{N}g_n(\hat\phi),
    \qquad
    \widetilde g_n
    =
    g_n(\hat\phi)-\bar g.
\]
The lag-$k$ covariance sum is defined as
\[
    \widehat{\Gamma}_k
    =
    \sum_{n=k+1}^{N}
    \widetilde g_n\widetilde g_{n-k}^{\top}.
\]
The resulting Newey-West estimator of the long-run score covariance
is
\begin{equation}
\label{eq:hacCovariance}
    J
    =
    \widehat{\Gamma}_0
    +
    \sum_{k=1}^{L}
    \left(1-\frac{k}{L+1}\right)
    \left(
        \widehat{\Gamma}_k
        +
        \widehat{\Gamma}_k^{\top}
    \right),
\end{equation}
where $L$ denotes the maximum lag included in the estimator.
The choice of truncation lag is an important component of HAC
estimation \citep{andrews1991}. In our implementation, we use
\[
    L
    =
    \min\left\{
        \left\lceil N^{1/3}\right\rceil,
        N-2
    \right\}.
\]

Combining the Hessian and the estimated long-run score covariance
yields the sandwich covariance estimator on the logarithmic
parameter scale:
\begin{equation}
\label{eq:sandwichCovariance}
    \widehat V_{\hat\phi}
    =
    \boldsymbol{H}^{-1}J\boldsymbol{H}^{-1}.
\end{equation}
Here, both $\boldsymbol{H}$ and $J$ are calculated using sums over the individual
observations, consistently with the definition of $\mathcal L$.
Consequently, no additional normalization by $N$ is required in
\eqref{eq:sandwichCovariance}.

This construction separates the curvature of the objective function
from the variability of its score contributions. In particular,
the HAC estimator allows the sandwich covariance to account for
possible temporal dependence between successive score contributions,
rather than relying on an independence assumption
\citep{neweyWest1987,andrews1991,zeileis2004}.

We construct confidence intervals using the asymptotic normal
approximation on the logarithmic parameter scale. Specifically,
for a confidence level $1-\alpha$, the marginal confidence interval
for the $j$th parameter is given by
\begin{equation}
\label{eq:sandwichCI}
    \left[
        \exp\left(
            \hat\phi_j
            -
            z_{1-\alpha/2}
            \sqrt{\widehat V_{\hat\phi,jj}}
        \right),
        \;
        \exp\left(
            \hat\phi_j
            +
            z_{1-\alpha/2}
            \sqrt{\widehat V_{\hat\phi,jj}}
        \right)
    \right],
\end{equation}
where $z_{1-\alpha/2}$ denotes the corresponding standard normal
quantile. We use $\alpha=0.05$ to obtain $95\%$ confidence
intervals. The exponential transformation ensures that the
resulting intervals remain positive and generally yields
asymmetric intervals on the original parameter scale.

These intervals are based on an asymptotic normal approximation
\citep{vanderVaart1998} and should therefore be interpreted as
approximate uncertainty measures. In particular, their accuracy
depends on the quality of the EKF-based approximate likelihood,
the numerical approximations of the Hessian and score contributions,
and the HAC estimation of their long-run covariance. Consequently,
nominal coverage is not guaranteed in finite samples. We therefore
assess the empirical coverage of the resulting sandwich-based
confidence intervals in the simulation study presented in
Section~\ref{sec:simulation}.

\begin{remark}
    In the following, we obtain the discrete-time state vector $\{X_n\}_{n=0}^N$ by evaluating the solution of a stochastic differential equation at the observation times via a numerical approximation method. Specifically, we consider the stochastic Jansen–Rit neural mass model and investigate its numerical approximation using the standard Euler–Maruyama method and two structure-preserving Strang splitting schemes. A central contribution of this work is incorporating these splitting methods into the EKF framework for parameter inference.
\end{remark}

\section{Stochastic Jansen-Rit neural mass model}
\label{sec:jr}

The Jansen–Rit neural mass model (JR-NMM), originally introduced by \citet{jansenRit}, is a well-established model for describing the average electrical activity of a population of neurons in the brain. In this work, we consider its stochastic extension proposed by \citet{haraldPaper}, which is formulated as a nonlinear six-dimensional stochastic differential equation (SDE).

In Section \ref{sec:jr:model}, we recall the formulation of the stochastic JR-NMM. Section \ref{sec:jr:hamiltonian} reformulates it as a stochastic Hamiltonian-type system. In addition, Section \ref{sec:jrNumerics} reviews numerical approximation methods for the model, namely the standard Euler--Maruyama method and the two structure-preserving Strang splitting schemes introduced in \citet{haraldPaper}.

\subsection{Model formulation}
\label{sec:jr:model}

\paragraph{Original model.}

In the original JR-NMM \citep{jansenRit}, a population of neurons, such as a cortical column, consists of three interacting groups: pyramidal cells, excitatory interneurons, and inhibitory interneurons. Pyramidal cells are the population's principal excitatory neurons, and they receive feedback from local excitatory and inhibitory interneurons, including nonpyramidal cells such as stellate and basket cells, respectively.

Two functional blocks describe each of these three groups. The first block converts the average incoming firing rate, i.e., the number of action potentials per unit time, into an average postsynaptic membrane potential, which may be excitatory or inhibitory. This transformation is modeled by a second-order ODE, which, for excitatory synaptic activity, takes the form
\begin{equation}
    \ddot{x}(t) = Aa\,z(t) - 2a\, \dot{x}(t) - a^2\,x(t),
\end{equation}
where $z(t)$ and $x(t)$ denote the input firing rate and output postsynaptic membrane potential, respectively. Here, $A>0$ is the average excitatory synaptic gain, while $a>0$ is the inverse time constant of the excitatory postsynaptic response. For inhibitory synaptic activity, the corresponding equation is
\begin{equation}
    \ddot{x}(t) = Bb\,z(t) - 2b\,\dot{x}(t) - b^2\,x(t),
\end{equation}
where $B>0$ denotes the average inhibitory synaptic gain and $b>0$ is the inverse time constant of the inhibitory postsynaptic response.

The second block converts the average membrane potential into an average firing rate. The sigmoid function models this nonlinear relationship
\begin{equation}
    \sigm(x)
    =
    \frac{\nu_{\max}}
         {1+\exp\left(r(\nu_0-x)\right)},
    \label{eq:sigmoid}
\end{equation}
where $\nu_{\max}$ denotes the maximum firing rate, $\nu_0$ is the membrane potential at which half of the maximum firing rate is attained, and $r$ determines the steepness of the sigmoid function.

Let $x_1(t)$, $x_2(t)$, and $x_3(t)$ denote the average postsynaptic membrane potentials associated with the pyramidal cells, excitatory interneurons, and inhibitory interneurons, respectively. Combining the two functional blocks yields the following system of three coupled second-order ODEs:
\begin{align}\label{eq:JR_original}
	\begin{split}
		\ddot{x}_1(t)&= Aa \, \sigm(x_2(t)-x_3(t))  -2a\,\dot{x}_1(t)-a^2\,x_1(t),\\
		\ddot{x}_2(t)&=Aa \left[ p(t)+ C_2 \, \sigm(C_1 x_1(t)) \right] -2a\,\dot{x}_2(t)-a^2\,x_2(t),\\
		\ddot{x}_3(t)&= Bb C_4 \, \sigm(C_3 x_1(t))  -2b\,\dot{x}_3(t)-b^2\,x_3(t),
	\end{split}	
\end{align}
where $C_1$, $C_2$, $C_3$, and $C_4$ are connectivity parameters describing the interactions between the pyramidal cells and the excitatory and inhibitory interneurons. Following the standard parametrization of the JR-NMM, these parameters are expressed in terms of a single connectivity parameter $C>0$, representing the average number of synaptic contacts between the neuronal groups. Table~\ref {tab:typicalValues} summarizes the physiological interpretation and typical values of the model parameters. Moreover, $p(t)$ is an external input function representing excitatory activity originating from neighboring or more distant regions of the brain. 

The average membrane potential received by the pyramidal cells, given by $y(t)=x_2(t)-x_3(t)$, represents the model's EEG-related output.

\begin{table}[t]
    \centering
    \begin{tabular}{|c|c|c|}
        \hline
        Parameter & Typical Value&Meaning \\
        \hline
        $C$ & $135$& Average number of synapses between the subpopulations \\
        $C_1,C_2$ & $C,\;0.8C$ & Avg. number of synapses in the excitatory loop\\
        $C_3,C_4$ & $0.25C,\;0.25C$ & Avg. number of synapses in the inhibitory loop \\
        $\nu_{\max}$ & $5s^{-1}$& Maximum firing rate of neurons \\
        $\nu_0$ & $6mV$& Value for which 50\% of the firing rate is attained \\
        $r$ & $0.56mV^{-1}$ & Slope of the sigmoid \\
        $A$ & $3.25mV$ & Average excitatory synaptic gain \\
        $a$ & $100ms^{-1}$ & Time constant of excitatory postsynaptic potential \\
        $B$ & $22mV$ & Average inhibitory synaptic gain \\
        $b$ & $50ms^{-1}$ & Time constant of inhibitory postsynaptic potential \\
        \hline
    \end{tabular}
    \caption{Typical parameter values for the stochastic JR-NMM.}
    \label{tab:typicalValues}
\end{table}

\paragraph{Stochastic model.}

To account for the stochastic nature of the external input $p(t)$, \citet{haraldPaper} formulated the JR-NMM as an SDE. Specifically, introducing the membrane potentials as position variables and their derivatives as velocity variables, the resulting stochastic JR-NMM model is described by the following six-dimensional nonlinear SDE:
\begin{align}
    dX^1(t)
    &= X^4(t)\,dt, \nonumber\\
    dX^2(t)
    &= X^5(t)\,dt, \nonumber\\
    dX^3(t)
    &= X^6(t)\,dt, \nonumber\\
    dX^4(t)
    &=
    \left[
        Aa\,\sigm(X^2(t)-X^3(t))
        -2a\,X^4(t)-a^2\,X^1(t)
    \right]dt
    +\tau\,dW^4(t), \nonumber\\
    dX^5(t)
    &=
    \left[
        Aa\left(
            \mu+C_2\,\sigm(C_1X^1(t))
        \right)
        -2a\,X^5(t)-a^2\,X^2(t)
    \right]dt
    +\sigma\,dW^5(t), \nonumber\\
    dX^6(t)
    &=
    \left[
        BbC_4\,\sigm(C_3X^1(t))
        -2b\,X^6(t)-b^2\,X^3(t)
    \right]dt
    +\tau\,dW^6(t),
    \label{eq:jrSDE}
\end{align}
where $t \in [0,T]$, 
\[
    W(t) =
    \left(
        W^4(t),W^5(t),W^6(t)
    \right)^\top, \qquad t \in [0,T],
\]
is a three-dimensional standard Brownian motion, and $X(0)=X_0$ is the initial contition. The parameters $\mu$ and $\sigma>0$ describe the mean and intensity of the external excitatory input, respectively, replacing the input function $p(t)$ in the original model. Additional stochastic perturbations act on the remaining two velocity components with intensity $\tau>0$, typically assumed to be much smaller than $\sigma$. Throughout this paper, we set $\tau=1$.

The SDE formulation also enables rigorous mathematical analysis of the stochastic JR-NMM. In particular, \citet{haraldPaper} established the existence and uniqueness of a strong solution 
\begin{equation*}
    X(t)=\left(X^1(t),\ldots,X^6(t)\right)^\top, \qquad t\in [0,T],
\end{equation*} 
and proved important structural properties of the model, including hypoellipticity and geometric ergodicity. In particular, hypoellipticity holds despite the degenerate diffusion, as the drift propagates noise to components not directly affected by the stochastic forcing.  

The model output corresponding to the EEG signal is given by the one-dimensional process
\begin{equation}
    Y(t) = X^2(t)-X^3(t), \qquad t \in [0,T].
    \label{eq:jrObservation}
\end{equation}
Thus, the six-dimensional state process $X(t)$  
generates an EEG-like signal through the difference between the excitatory and inhibitory postsynaptic membrane potentials received by the pyramidal cells.

\begin{figure}
    \centering
    \includegraphics[width=1.0\linewidth]{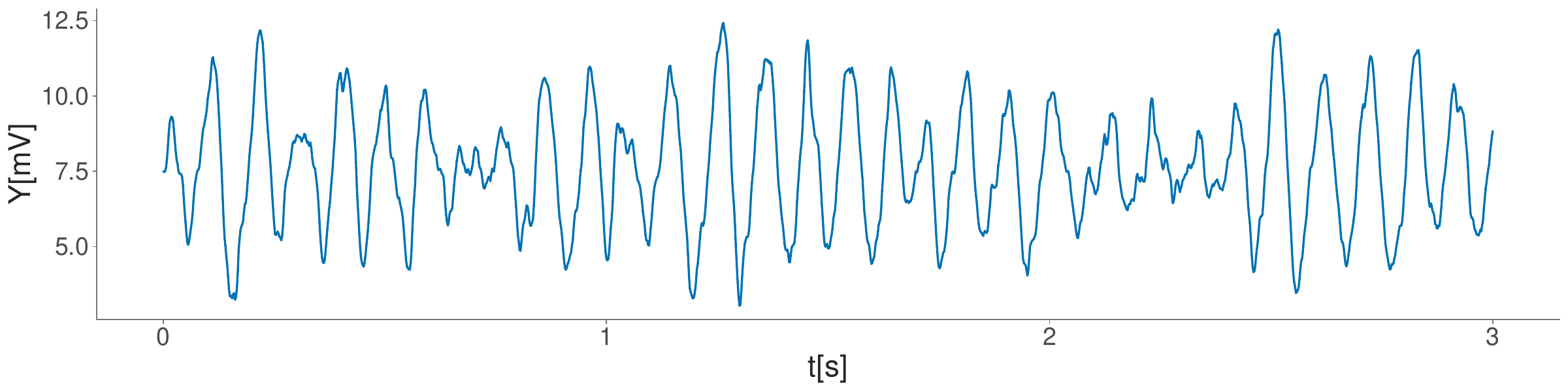}
    \caption{Sample trajectory using parameters obtained form Table \ref{tab:typicalValues}.}
    \label{fig:alphaRhythm}
\end{figure}

Figure \ref{fig:alphaRhythm} shows a typical sample path of $Y(t)$, simulated using the parameter values in Table~\ref{tab:typicalValues} and the Ornstein--Uhlenbeck Strang splitting method described in Section~\ref{sec:jrNumerics:splitting}~below. In particular, the standard connectivity value $C=135$ corresponds to an \textit{alpha rhythm}, characterized by oscillations in the frequency range of approximately $8$--$13$ Hz. Varying $C$ can substantially change the model's oscillatory behavior and lead to qualitatively different output signals.

In this paper, we focus on the simultaneous inference of the parameter vector
\begin{equation}
\label{eq:theta}
     \theta=(C,\mu,\sigma).
\end{equation}
The connectivity parameter $C$ already appears in the original JR-NMM and strongly influences its dynamics. The parameters $\mu$ and $\sigma$, introduced in the stochastic extension, govern the mean and variability of the external excitatory input and likewise substantially influence the model output. These three parameters are therefore of particular interest for parameter inference. To the best of our knowledge, only \cite{ireneSpringer} has considered their simultaneous estimation using approximate Bayesian computation (ABC). Here, we investigate their joint estimation using approximate likelihoods based on the extended Kalman~filter.

For details on the stochastic JR-NMM, its mathematical properties, and its physiological interpretation, see \cite{haraldPaper} and the references therein.

\subsection{Reformulation as a stochastic Hamiltonian-type system}
\label{sec:jr:hamiltonian}

Following \cite{haraldPaper}, we reformulate the stochastic JR-NMM \eqref{eq:jrSDE} as a stochastic Hamiltonian-type system. This representation provides a useful basis for the numerical approximation methods discussed in Section \ref{sec:jrNumerics}, which we incorporate into the EKF framework for parameter inference in Section \ref{sec:jrEkf}. 

To this end, we introduce the position and
velocity variables
\[
    Q(t) =
    \left(
        X^1(t),X^2(t),X^3(t)
    \right)^\top,
    \qquad
    P(t) =
    \left(
        X^4(t),X^5(t),X^6(t)
    \right)^\top, \qquad t \in [0,T],
\]
and set 
\begin{equation*}
   X(t) =
    \begin{pmatrix}
        Q(t)\\
        P(t)
    \end{pmatrix}, \qquad t \in [0,T].
\end{equation*}
Moreover, we define the diagonal matrices
\[
    \Gamma
    =
    \operatorname{diag}(a,a,b),
    \qquad
    \Sigma
    =
    \operatorname{diag}(\tau,\sigma,\tau).
\]
The drift of SDE \eqref{eq:jrSDE} decomposes into three components: a Hamiltonian part, a linear damping term, and a nonlinear forcing term. The Hamiltonian part is associated with the function
\begin{equation}
    \mathcal{H}(Q,P)
    =
    \frac12
    \left(
        \|P\|_{\mathbb R^3}^2
        +
        \|\Gamma Q\|_{\mathbb R^3}^2
    \right),
    \label{eq:jrHamiltonian}
\end{equation}
for which
\[
    \nabla_P \mathcal{H}(Q,P)=P,
    \qquad
    \nabla_Q \mathcal{H}(Q,P)=\Gamma^2Q.
\]
The linear damping term is given by $-2\Gamma P$, while the nonlinear forcing term is represented by the nonlinear function
\begin{equation}
    G(Q)
    =
    \begin{pmatrix}
        Aa\,\sigm(Q^2-Q^3)\\[3pt]
        Aa\left(
            \mu+C_2\sigm(C_1Q^1)
        \right)\\[3pt]
        BbC_4\sigm(C_3Q^1)
    \end{pmatrix}.
    \label{eq:jrNonlinearity}
\end{equation}

Combining these three components, the stochastic JR-NMM \eqref{eq:jrSDE} can be written compactly as
\begin{equation}
\label{eq:jrHamiltonianForm}
    d\begin{pmatrix}
        Q(t)\\[2pt]
        P(t)
    \end{pmatrix}
    =
    \begin{pmatrix}
        \nabla_P \mathcal{H}(Q(t),P(t))\\[2pt]
        -\nabla_Q \mathcal{H}(Q(t),P(t))
        -2\Gamma P(t)
        +G(Q(t))
    \end{pmatrix}dt
    +
    \begin{pmatrix}
        \mathbb{O}_3\\[2pt]
        \Sigma
    \end{pmatrix}dW(t).
\end{equation}

Alternatively, introducing
\begin{equation}
    M
    =
    \begin{pmatrix}
        \mathbb O_{3} &\mathbb I_3\\
        -\Gamma^2 & -2\Gamma
    \end{pmatrix},
    \qquad
    G_0(X) = \begin{pmatrix}
        0_3\\G(Q)
    \end{pmatrix},
    \qquad
    \Sigma_0
    =
    \begin{pmatrix}
        \mathbb O_{3}\\
        \Sigma
    \end{pmatrix},
    \label{eq:jrDecomposition}
\end{equation}
we obtain the semi-linear additive-noise SDE representation
\begin{equation}
\label{eq:additiveJR}
    dX(t)
    =
    \left(
        MX(t)+G_0(X(t))
    \right)dt
    +
    \Sigma_0\,dW(t),
\end{equation}
which is equivalent to \eqref{eq:jrHamiltonianForm}. 

The structure of these representations can be used in the numerical approximation methods discussed below.

\subsection{Numerical approximation methods}
\label{sec:jrNumerics}

Since the solution of SDE \eqref{eq:jrSDE} (equivalently, \eqref{eq:jrHamiltonianForm} and \eqref{eq:additiveJR}) is not available,  numerical methods are required to approximate it. Consider an equidistant discretisation $0=t_0<t_1<\ldots<t_N=T$ of the time interval $[0,T]$, with step size $\Delta=t_n-t_{n-1}$, $n=1,\ldots,N$, so that $t_n=n\Delta$ for $n=0,\ldots,N$. We seek numerical approximations 
\begin{equation*}
    {X}_{n} \approx X({t_n})
\end{equation*}
of the solution at these discrete time points. Specifically, we consider the standard Euler--Maruyama method \citep{KloedenPlaten} and two Strang splitting methods, namely the Wiener and Ornstein--Uhlenbeck integrators, introduced in \cite{haraldPaper}.

\subsubsection{Euler--Maruyama method}

The Euler--Maruyama method is perhaps the most widely used numerical method for SDEs (see, e.g., \cite{KloedenPlaten}). It is obtained by truncating the stochastic Taylor expansion after the leading-order terms. Applying the Euler--Maruyama method directly to SDE \eqref{eq:jrHamiltonianForm}, we obtain the following iteration for $n=1,\ldots,N$, initialized with ${X}^{\textrm{EM}}_{0}=X_0$:
\begin{equation}
    \begin{aligned}
\label{eq:oneStepEuler}
     X^{\textrm{EM}}_{n}
    &=
     X^{\textrm{EM}}_{{n-1}}
    +
    \Delta
    \begin{pmatrix}
        P^{\textrm{EM}}_{{n-1}}\\[0.1cm]
        -\Gamma^2 Q^{\textrm{EM}}_{{n-1}}
        -2\Gamma P^{\textrm{EM}}_{{n-1}}
        +G( Q^{\textrm{EM}}_{{n-1}})
    \end{pmatrix}
    + \underbrace{\Sigma_0 \left( W({t_n}) - W({t_{n-1}}) \right)}_{=:\xi_{n,\Delta}} \\
    &=f^{\textrm{EM}}_\Delta( X^{\textrm{EM}}_{n-1}, \xi_{n,\Delta} ).
\end{aligned}
\end{equation}
Here, $\xi_{n,\Delta}$, $n=1,\ldots,N$, are independent and identically distributed Gaussian random vectors with
\begin{equation}
\label{eq:EM_xi}
    \xi_{n,\Delta}\sim\mathcal N\left(0_6,\mathcal C_\Delta \right), \quad \mathcal C_\Delta = \Delta \Sigma_0\Sigma_0^\top = \Delta \begin{pmatrix}
    \mathbb{O}_3 & \mathbb{O}_3 \\
    \mathbb{O}_3 & \Sigma^2
    \end{pmatrix}.    
\end{equation}

Note that the transition distribution of the Euler--Maruyama method applied to the stochastic JR-NMM, i.e., the conditional distribution of $ X^{\textrm{EM}}_{n}$ given $ X^{\textrm{EM}}_{{n-1}}$, is a degenerate Gaussian distribution, as its covariance matrix $\mathcal C_\Delta$ only has rank three. 

\subsubsection{Strang splitting methods}
\label{sec:jrNumerics:splitting}

The numerical splitting approach differs fundamentally from the standard Euler--Maruyama method. It consists of three main steps:
\begin{itemize}
    \item[i.] Split the original (unsolvable) SDE into subequations that can be solved exactly.
    \item[ii.] Determine the exact solutions of these subequations.
    \item[iii.] Compose the corresponding solution maps in a suitable way to obtain a numerical approximation of the original SDE over each time step. 
\end{itemize}
Here, we consider the two Strang splitting methods proposed in \cite{haraldPaper} for the stochastic JR-NMM, which use a symmetric composition of the solution maps.

\paragraph{Wiener Strang splitting.}

We first recall the Wiener Strang splitting method introduced in \cite{haraldPaper}.

\textit{Step i.} The dynamics of SDE \eqref{eq:additiveJR} are decomposed into a deterministic linear subsystem and a stochastic nonlinear subsystem:
\begin{align}
dX^{[1]}(t)
&= M X^{[1]}(t) \, dt, \qquad X^{[1]}(0)=X_0^{[1]},
\label{eq:wienerFlow1}\\
dX^{[2]}(t)
&= G_0(X^{[2]}(t)) \, dt
+ \Sigma_0 dW(t), \qquad X^{[2]}(0)=X_0^{[2]}.
\label{eq:wienerFlow2}
\end{align}

\textit{Step ii.} The first subequation is a linear ODE with the exact solution
\begin{equation}
X^{[1]}(t) = \psi_t^{[1]}(X_0^{[1]})
=
e^{M t}X_0^{[1]}.
\label{eq:wienerSolution1}
\end{equation}
Due to the particular block structure of the matrix $M$ in \eqref{eq:jrDecomposition} and the diagonal form of $\Gamma$, the matrix exponential can be computed explicitly as
\begin{equation}
\label{eq:matrixExponential}
    e^{M t}
    =
    \begin{pmatrix}
        e^{-\Gamma t}(\mathbb I_3+\Gamma t)
        &
        t e^{-\Gamma t}
        \\[4pt]
        -\Gamma^2 t e^{-\Gamma t}
        &
        e^{-\Gamma t}(\mathbb I_3-\Gamma t)
    \end{pmatrix} =: \begin{pmatrix}
        \vartheta_t&\kappa_t\\\vartheta'_t&\kappa'_t
    \end{pmatrix}.
\end{equation}

In the second subequation, the $Q$-component remains constant, while the $P$-component  depends only on this constant $Q$-component (cf. $G_0$ and $\Sigma_0$ in \eqref{eq:jrDecomposition}). Hence, this subequation can also be solved exactly, yielding
\begin{equation}
X^{[2]}(t) = \psi_t^{[2]}(X_0^{[2]})
= X_0^{[2]} + t G_0 (X_0^{[2]}) + \Sigma_0 W(t),
\label{eq:wienerSolution2}
\end{equation}
where 
\begin{equation*}
    \Sigma_0 W(t) \sim \mathcal N\left( 0_6, t \Sigma_0\Sigma_0^\top \right).
\end{equation*}

\textit{Step iii.} Composing the exact solution maps in \eqref{eq:wienerSolution1} and \eqref{eq:wienerSolution2} symmetrically over each time step yields the Wiener Strang splitting method, defined recursively by
\begin{equation}
\begin{aligned}
\label{eq:wienerStrang}
 X^{\textrm{WIE}}_{n} &= \left( \psi_{\Delta/2}^{[1]}
\circ
\psi_{\Delta}^{[2]}
\circ
\psi_{\Delta/2}^{[1]} \right) \left( X^{ \textrm{WIE}}_{{n-1}}\right) \\[0.1cm]
&=
e^{M\Delta}  X^{\textrm{WIE}}_{{n-1}} + \Delta e^{M \Delta/2} G_0\left( e^{M\Delta/2}  X^{\textrm{WIE}}_{{n-1}} \right) + \underbrace{e^{M\Delta/2} \Sigma_0 \left( W({t_n}) -W({t_{n-1}}) \right)}_{=:\xi_{n,\Delta}} \\
&=f^{\textrm{WIE}}_\Delta( X^{\textrm{WIE}}_{{n-1}},\xi_{n,\Delta}),
\end{aligned}
\end{equation}
for $i=1,\ldots,N$, with initial condition $ X^{\textrm{WIE}}_{0}=X_0$. Now, $\xi_{n,\Delta}$, $n=1,\ldots,N$, are independent and identically distributed Gaussian random vectors with
\begin{equation}
\label{eq:WIE_xi}
        \xi_{n,\Delta}\sim\mathcal N\left(0_6,\mathcal C_\Delta \right), \quad \mathcal C_\Delta = \Delta e^{M\Delta/2} \Sigma_0\Sigma_0^\top \left( e^{M\Delta/2} \right)^\top
    = \Delta \begin{pmatrix}
        \kappa_{\Delta/2} \Sigma^2 \kappa_{\Delta/2}^\top & \kappa_{\Delta/2} \Sigma^2 \kappa_{\Delta/2}^{\prime\top} \\[0.2cm]
        \kappa_{\Delta/2}' \Sigma^2 \kappa_{\Delta/2}^\top & \kappa_{\Delta/2}' \Sigma^2 \kappa_{\Delta/2}^{\prime\top}
    \end{pmatrix}.
\end{equation}

Since $\kappa_{\Delta/2}$, $\kappa_{\Delta/2}'$, and $\Sigma$ are diagonal matrices, and $\kappa_{\Delta/2}$ is invertible for $\Delta>0$, the lower three rows of $\mathcal C_\Delta$ are obtained by multiplying the upper three rows by $\kappa_{\Delta/2}' \kappa_{\Delta/2}^{-1}$. Thus, $\mathcal C_\Delta$ has rank three. Consequently, the transition distribution of the Wiener Strang splitting method, i.e., the conditional distribution of $X^{\textrm{WIE}}_{{n}}$ given $X^{\textrm{WIE}}_{{n-1}}$, is degenerate Gaussian. Nevertheless, unlike the Euler--Maruyama method, the covariance matrix of the Wiener Strang splitting method has no zero entries. This is because of the matrix exponential $e^{M\Delta/2}$, which accounts for the interaction between the position and velocity components.

\paragraph{Ornstein--Uhlenbeck Strang splitting.}

We now consider the Ornstein--Uhlenbeck Strang splitting method introduced by \cite{haraldPaper}.

\textit{Step i.} In this method, SDE \eqref{eq:additiveJR} is decomposed into a linear stochastic subsystem and a nonlinear deterministic subsystem:
\begin{align}
    dX^{[1]}(t)
    &=
    MX^{[1]}(t)\,dt
    +
    \Sigma_0\,dW(t), \qquad X^{[1]}(0)=X_0^{[1]},
    \label{eq:ouFlow1}\\
    dX^{[2]}(t)
    &= G_0(X^{[2]}(t))\,dt, \qquad X^{[2]}(0)=X_0^{[2]}.  
    \label{eq:ouFlow2}
\end{align}

\textit{Step ii.} The first subequation is a multivariate Ornstein--Uhlenbeck process with the exact solution
\begin{equation}
\label{eq:ouTransition}
    X^{[1]}(t) = \psi_t^{[1]}(X_0^{[1]}) = e^{Mt}X_0^{[1]} + \underbrace{\int_0^t e^{M(t-s)} \Sigma_0 \,dW_s}_{=: \xi_t},
\end{equation}
where the matrix exponential $e^{Mt}$ is as in \eqref{eq:matrixExponential} and
\begin{equation*}
    \xi_t \sim \mathcal{N}\left( 0_6, C_t \right), \quad \mathcal C_t = \int_0^t e^{M(t-s)} \Sigma_0\Sigma_0^\top \left( e^{M(t-s)} \right)^\top \, ds.
\end{equation*}
Due to the particular block structure of $M$ in \eqref{eq:jrDecomposition}, arising from the Hamiltonian formulation of the stochastic JR-NMM, and the diagonal form of $\Gamma$, the covariance matrix $\mathcal C_t$ can also be computed explicitly, with its closed-form expression involving the blocks of the matrix exponential $e^{Mt}$ derived in \eqref{eq:matrixExponential}. Specifically, we obtain
\begin{equation}
    \mathcal C_t
     = \begin{pmatrix}
        \frac14\Gamma^{-3}\Sigma^2(\mathbb I_3+\kappa_t\vartheta'_t-\vartheta_t^2)&
        \frac12\Sigma^2\kappa^2_t\\[0.2cm] \frac12\Sigma^2\kappa^2_t&
        \frac14\Gamma^{-1}\Sigma^2(\mathbb I_3+\kappa_t\vartheta'_t-\kappa_t'^{2})
    \end{pmatrix}.
    \label{eq:ouCovIntegral}
\end{equation}

In the second subequation, similar to before, the $Q$-component remains constant, while the $P$-component depends only on this constant $Q$-component. This subequation can thus also be solved exactly, yielding 
\begin{equation}
\label{eq:nonlinearFlowJR}
    X^{[2]}(t) = \psi_t^{[2]}(X_0^{[2]})=X_0^{[2]} + t G_0(X_0^{[2]}).
\end{equation}

\textit{Step iii.} Composing the exact solution maps in \eqref{eq:ouTransition} and \eqref{eq:nonlinearFlowJR} symmetrically over each time step yields the Ornstein--Uhlenbeck Strang splitting method, defined recursively by 
\begin{equation}
    \begin{aligned}\label{eq:ouStrangTransition}
     X^{\textrm{OU}}_{n} &= \left( \psi_{\Delta/2}^{[2]}
\circ
\psi_{\Delta}^{[1]}
\circ
\psi_{\Delta/2}^{[2]} \right) \left( X^{\textrm{OU}}_{{n-1}}\right) \\[0.1cm]
&= e^{M\Delta} X^{\textrm{OU}}_{{n-1}} + \Delta/2 e^{M\Delta}G_0\left( X^{\textrm{OU}}_{{n-1}} \right) + \xi_{n,\Delta} \\[0.1cm] 
& \qquad + \Delta/2 G_0\left( e^{M\Delta} X^{\textrm{OU}}_{{n-1}} + \Delta/2 e^{M\Delta}G_0\left( X^{\textrm{OU}}_{{n-1}} \right) + \xi_{n,\Delta} \right) \\[0.1cm]
&=f^{\textrm{OU}}_\Delta(X^{\textrm{OU}}_{{n-1}},\xi_{n,\Delta}),
\end{aligned}
\end{equation}
for $i=1,\ldots,N$, with initial condition ${X}_{0}^{\textrm{OU}}=X_0$. Here, $\xi_{n,\Delta}$, $n=1,\ldots,N$, are again independent and identically distributed Gaussian random vectors with
\begin{equation}
\label{eq:OU_xi}
    \xi_{n,\Delta} \sim \mathcal{N}\left( 0_6, \mathcal C_\Delta \right),
\end{equation}
where $\mathcal C_\Delta$ is as in \eqref{eq:ouCovIntegral}. 

In contrast to the Euler--Maruyama and Wiener Strang splitting methods, the Gaussian random vectors $\xi_{n,\Delta}$ enter the nonlinear component $G_0$ of the stochastic JR-NMM in the Ornstein--Uhlenbeck Strang splitting method \eqref{eq:ouStrangTransition}. Consequently, the transition distribution of this numerical method is non-Gaussian. Moreover, unlike the other two methods, the covariance matrix $\mathcal C_\Delta$ of the $\xi_{n,\Delta}$ has full rank. 

\begin{remark}
    As shown by \citet{haraldPaper}, both splitting methods preserve important structural properties of the stochastic JR-NMM, including geometric ergodicity and moment bounds. In contrast, the Euler--Maruyama method does generally not retain such properties. These results suggest that splitting methods may also offer advantages for EKF-based parameter inference. 
    
    In addition, the numerical methods differ in their transition distributions, as discussed above. While the Euler--Maruyama method has a degenerate covariance matrix with zero position-related blocks, the Wiener Strang splitting method accounts for the interaction between position and velocity components, although its covariance matrix remains singular. In contrast, the Ornstein--Uhlenbeck Strang splitting method yields a full-rank covariance matrix for the Gaussian random vectors~$\xi_{n,\Delta}$. 

    Furthermore, the dependence of the nonlinear component $G_0$ on the $\xi_{n,\Delta}$ in the Ornstein--Uhlenbeck Strang splitting method affects the Jacobians used in the EKF (in particular, leading to a non-trivial Jacobian $F_{\xi,n}$, unlike in the other two methods) and, consequently, the approximate likelihood employed for parameter inference, as discussed in Section~\ref{sec:jrEkf}. 
    
    Taken together, these differences motivate investigating whether the splitting methods, particularly the Ornstein--Uhlenbeck method, improve EKF-based parameter inference compared with Euler--Maruyama.
\end{remark}

\begin{remark}
    Note that the order of composition in \eqref{eq:wienerStrang} and \eqref{eq:ouStrangTransition}, respectively, could also be reversed. Here, we follow \citet{haraldPaper} and choose the composition such that the deterministic subequation is solved over two half-steps, while the stochastic subequation is solved over one full step. This is computationally advantageous, as the reversed composition would require generating two stochastic increments per time step instead of one.

    An alternative to Strang splitting is the Lie–Trotter splitting method, which \citet{haraldPaper} also considers. However, their numerical results indicate that Strang splitting generally performs better. This may be due to its symmetric composition \citep{haraldPaper} or higher-order one-step predictions \citep{pilipovic2024}.
\end{remark}

\section{Extended Kalman filter-based parameter inference for the stochastic JR-NMM}
\label{sec:jrEkf}

The numerical approximation methods introduced in Section~\ref{sec:jr} provide the state transition functions required to apply the EKF framework of Section~\ref{sec:ekf} to the stochastic JR-NMM. Each method defines a discrete-time state transition of the form
\[
X_n=f_\Delta(X_{n-1},\xi_{n,\Delta}),
\qquad n=1,\ldots,N,
\]
which approximates the evolution of the continuous-time model over the interval $[t_{n-1},t_n]$ of length $\Delta$.

The choice of the transition function $f_\Delta$ is important for parameter inference, as it determines the predicted state moments and the Jacobians used in the EKF recursion and, consequently, the approximate innovation likelihood. We therefore investigate the Euler--Maruyama, Wiener Strang splitting, and Ornstein--Uhlenbeck Strang splitting methods in this setting. While \citet {haraldDiss} previously derived the EKF formulations for the Euler--Maruyama and Wiener Strang splitting methods, we extend the analysis to the Ornstein--Uhlenbeck Strang splitting method and consider all three methods in the context of parameter inference, which that work did not address. In particular, we derive the Jacobians required for the Ornstein--Uhlenbeck Strang splitting method, whose nonlinear dependence on the process noise leads to a non-trivial noise Jacobian $F_{\xi,n}$, unlike in the other two methods.

In Section~\ref{sec:transFunc_Jacobians}, we derive the Jacobians of the numerical state transition functions required by the EKF. Section~\ref{sec:obs_model} introduces the observation model and its Jacobians, while Section~\ref{sec:jrParameterLikelihood} describes the resulting EKF-based parameter inference methods. Finally, Section~\ref{sec:substepping} presents a time-step refinement procedure for cases where observation intervals are too large for sufficiently accurate numerical approximation of the state dynamics.

\subsection{Transition functions and Jacobi matrices}
\label{sec:transFunc_Jacobians}

In this section, we specify the numerical transition functions and derive their Jacobians for the~EKF. For completeness, we first recall the Jacobians for the
Euler--Maruyama and Wiener Strang splitting methods derived by \cite{haraldDiss}, before deriving those for the Ornstein--Uhlenbeck Strang splitting method.

\subsubsection{Euler--Maruyama transitions}

Let $x=(q,p)^\top \in \mathbb{R}^6$ denote an arbitrary state vector and let $\xi \in \mathbb{R}^6$ represent the Gaussian random vector $\xi_{n,\Delta}$ defined in~\eqref{eq:EM_xi}. Then the Euler--Maruyama method \eqref{eq:oneStepEuler} yields the following transition function for the~EKF:
\begin{equation}
\label{eq:emTransitionEKF}
    f_{\Delta}^{\mathrm{EM}}(x,\xi)
    =
    x
    +
    \Delta
    \begin{pmatrix}
        p\\
        -\Gamma^2q-2\Gamma p+G(q)
    \end{pmatrix}
    + \xi.
\end{equation}

Its state and noise Jacobians are therefore given by
\begin{equation}
\label{eq:emJacobians}
    F_x^{\mathrm{EM}}(x)
    =
    \mathbb I_6+
    \Delta
    \begin{pmatrix}
        \mathbb{O}_3&\mathbb I_3\\
        -\Gamma^2+G_q(q)&-2\Gamma
    \end{pmatrix},
    \qquad
    F_\xi^{\mathrm{EM}}
    = \mathbb{I}_6,
\end{equation}
where $G_q$ denotes the Jacobian of the nonlinear forcing term 
\eqref{eq:jrNonlinearity}. Specifically, it is given by
\begin{equation}
\label{eq:GQ}
    G_q(q) = \frac{\partial G(q)}{\partial q}
    =
    \begin{pmatrix}
        0
        &
        Aa\,\sigm'(q^2-q^3)
        &
        -Aa\,\sigm'(q^2-q^3)
        \\[3pt]
        AaC_1C_2\,\sigm'(C_1q^1)
        &
        0
        &
        0
        \\[3pt]
        BbC_3C_4\,\sigm'(C_3q^1)
        &
        0
        &
        0
    \end{pmatrix},
\end{equation}
where, for the sigmoid function defined in \eqref{eq:sigmoid}, one has
\begin{equation}
\label{eq:sigmoidDerivative}
    \sigm'(x)
    =
    r\,\sigm(x)
    \left(
        1-\frac{\sigm(x)}{\nu_{\max}}
    \right).
\end{equation}
Note that, since the process noise $\xi$ enters
the transition function $f_\Delta^{\textrm{EM}}$ additively, its Jacobian $F_\xi^{\textrm{EM}}$ is independent of the current
state $x$.

\subsubsection{Wiener Strang splitting transitions}

Let $x=(q,p)^\top \in \mathbb{R}^6$ denote an arbitrary state vector and let $\xi \in \mathbb{R}^6$ represent the Gaussian random vector $\xi_{n,\Delta}$ defined in \eqref{eq:WIE_xi}. Then the Wiener Strang splitting method \eqref{eq:wienerStrang} yields the following transition function for the EKF:
\[
f_\Delta^{\textrm{WIE}}(x,\xi)=e^{M\Delta}x+\Delta e^{M\Delta/2} G_0\left( e^{M\Delta/2} x \right) + \xi.
\]

As in the Euler--Maruyama case, the noise enters additively, so the noise Jacobian is given by
\begin{equation*}
    F_\xi^{\textrm{WIE}}= \mathbb I_6.
\end{equation*}

To obtain the state Jacobian, we first derive the Jacobian of the function $G_0$ defined in \eqref{eq:jrDecomposition}. Since $G_0(x)=(0_3,G(q))^\top$, we have
\begin{equation}
\label{eq:Gzero_x}
    (G_0)_x(x)=\frac{\partial G_0(x)}{\partial x} 
    =\begin{pmatrix}
    \mathbb O_{3}&\mathbb O_{3}\\G_q(q)&\mathbb O_{3}
    \end{pmatrix},
\end{equation}
where $G_q(q)$ is as in \eqref{eq:GQ}. Applying the chain rule then yields
\[
F_x^{\textrm{WIE}}(x) = e^{M\Delta}+ \Delta e^{M\Delta/2} (G_0)_x\left(e^{M\Delta/2}x\right)e^{M\Delta/2}. 
\]

\subsubsection{Ornstein--Uhlenbeck Strang splitting transitions}
\label{sec:ouEkf}

Let $x=(q,p)^\top \in \mathbb{R}^6$ denote an arbitrary state vector and let $\xi \in \mathbb{R}^6$ represent the Gaussian random vector $\xi_{n,\Delta}$ of the Ornstein--Uhlenbeck flow defined in \eqref{eq:OU_xi}. Recall that one step of the Ornstein--Uhlenbeck Strang splitting method
\eqref{eq:ouStrangTransition} consists of a deterministic nonlinear
half-step, a stochastic linear full step, and a second deterministic nonlinear
half-step. 
The corresponding transition function for the EKF can therefore be written as
\begin{equation}
\label{eq:ouCompositionCompact}
    f_\Delta^{\textrm{OU}}(x,\xi)
    =
    \psi_{\Delta/2}^{[2]}
    \left(
        e^{M\Delta}
        \psi_{\Delta/2}^{[2]}(x)
        +
        \xi
    \right),
\end{equation}
where $\psi^{[2]}$ is defined in \eqref{eq:nonlinearFlowJR}.

To derive the state and noise Jacobians, we first consider the Jacobian of the function $\psi_{\Delta/2}^{[2]}$ for the nonlinear half-step. Specifically, we obtain
\begin{equation}
\label{eq:BhalfExplicit}
    B_{\Delta/2}(x)
    :=
    \frac{\partial\psi_{\Delta/2}^{[2]}(x)}
         {\partial x} =
    \mathbb{I}_6 + \frac{\Delta}{2}(G_0)_x(x) =
    \begin{pmatrix}
        \mathbb I_3& \mathbb{O}_3\\[2pt]
        \dfrac{\Delta}{2}G_q(q)&\mathbb I_3
    \end{pmatrix},
\end{equation}
where $(G_0)_x(x)$ is as in \eqref{eq:Gzero_x} and $G_q(q)$ is given in \eqref{eq:GQ}. The Jacobian matrix in \eqref{eq:BhalfExplicit} accounts for the nonlinear half-steps in the Strang composition and will be used to derive the Jacobians of the full transition function.

Applying the chain rule to \eqref{eq:ouCompositionCompact} using $B_{\Delta/2}$ from~\eqref{eq:BhalfExplicit}, we obtain the state Jacobian 
\begin{align}
\label{eq:FxOU}
    F_x^{\mathrm{OU}}(x,\xi)
    =B_{\Delta/2}
    \left(
        e^{M\Delta}\psi_{\Delta/2}^{[2]}(x)+\xi
    \right)
    e^{M\Delta}
    B_{\Delta/2}(x).
\end{align}

Similarly, differentiating \eqref{eq:ouCompositionCompact} with respect
to $\xi$ yields the noise Jacobian 
\begin{equation}
\label{eq:FxiOU}
    F_{\xi}^{\mathrm{OU}}(x,\xi)
    =
    B_{\Delta/2}
    \left(
        e^{M\Delta}\psi_{\Delta/2}^{[2]}(x)+\xi
    \right).
\end{equation}

Unlike the Euler--Maruyama and Wiener Strang splitting methods, the Ornstein--Uhlenbeck Strang splitting method yields a non-trivial noise Jacobian $F^{\textrm{OU}}_\xi$, which depends on both the state $x$ and the process noise $\xi$. When evaluated at the EKF linearization point $(x,\xi)=(m_{n-1},0_6)$, this Jacobian, together with the non-degenerate process noise covariance matrix $\mathcal{C}_\Delta$, affects the propagation of process noise in the predicted state covariance (cf. the second term in \eqref{eq:ekfPredictionCov}) and, consequently, the innovation covariance entering the approximate likelihood.

\subsection{Observation model}
\label{sec:obs_model}

Recall that the EEG-like output of the stochastic JR-NMM is given by $Y(t)=X^2(t)-X^3(t)$, $t \in [0,T]$, as introduced in \eqref{eq:jrObservation}. At the discrete observation times $t_n$, $n=1,\ldots,N$, we consider noisy observations of this signal given by 
\begin{equation}
\label{eq:jrObservationEKF}
    Y_n
    =
    X_n^2-X_n^3+r_n,
    \qquad
    r_n\sim\mathcal N(0,R),
\end{equation}
where $R\geq 0$ denotes the observation-noise variance.

Let $x=(x^1,\ldots,x^6)^\top\in\mathbb R^6$ denote an arbitrary state vector and let $r \in \mathbb{R}$ represent the Gaussian observation noise $r_n$ defined in \eqref{eq:jrObservationEKF}. The corresponding observation function for the EKF is 
\begin{equation*}
    h(x,r)=x^2-x^3+r.
\end{equation*}
Its state and noise Jacobians are constant and given by
\begin{equation}
\label{eq:jrObservationJacobians}
    H_x
    =
    \begin{pmatrix}
        0&1&-1&0&0&0
    \end{pmatrix},
    \qquad
    H_r=1.
\end{equation}
Consequently, the innovation covariance in the EKF recursion (cf. \eqref{eq:innovationCov}) takes the form
\begin{equation}
\label{eq:jrInnovationCov}
    S_n
    =
    H_x\bar P_nH_x^\top+R.
\end{equation}
Since the observation function is linear, it introduces no additional linearization error in the EKF. All nonlinear filtering effects enter through the
state transition.

\subsection{EKF-based inference methods}
\label{sec:jrParameterLikelihood}

Combining the numerical state transition functions and their Jacobians from Section~\ref{sec:transFunc_Jacobians} with the observation model from Section \ref{sec:obs_model}, we obtain three EKF-based approaches to parameter inference for the stochastic JR-NMM. In the following, we refer to these methods as
\begin{itemize}
    \item \textbf{EM-EKF}, based on the Euler--Maruyama (EM) method,
    \item \textbf{WIE-EKF}, based on the Wiener Strang splitting (WIE) method,
    \item \textbf{OU-EKF}, based on the Ornstein--Uhlenbeck Strang splitting (OU) method.
\end{itemize}

For each method and each candidate parameter vector $\theta$, the EKF recursion in Section \ref{sec:ekf} yields the corresponding innovations $\nu_n(\theta)$ and innovation covariance matrices $S_n(\theta)$, $n=1,\ldots,N$. We use these to evaluate the approximate innovation log-likelihood \eqref{eq:quasiLikelihood}, which we maximize as in \eqref{eq:qMLE} to obtain the parameter estimates.

Although all three approaches follow the same EKF-based inference framework, they differ in their numerical state transitions, Jacobians, and process noise covariance matrices, all of which may depend on $\theta$. These differences lead to distinct approximate innovation likelihoods and, consequently, yield different parameter estimates.

\subsection{Time-step refinement}
\label{sec:substepping}

The accuracy of any numerical approximation method depends strongly on the chosen time-step size. If the observation interval $\Delta$ is too large, the resulting numerical discretization error may also affect the accuracy of the EKF-based parameter inference. To address this issue, we consider time-step refinement. While the observations are fixed on the original time grid, the numerical transitions used inside the EKF can be evaluated on a finer internal grid. This is a common procedure when the observation step-size is not sufficiently small to ensure adequate accuracy of a numerical scheme \citep{bellanger2012}. 

Let $n_{\mathrm{sub}}$ denote the number of internal substeps within one
observation interval $\Delta$ and define
\begin{equation}
\label{eq:internalStep}
    \delta
    =
    \frac{\Delta}{n_{\mathrm{sub}}}.
\end{equation}
Instead of applying a single numerical step of length $\Delta$, we perform $n_{\mathrm{sub}}$ successive steps of length~$\delta$. For the Ornstein--Uhlenbeck Strang splitting method \eqref{eq:ouStrangTransition}, for example, we define 
\begin{equation}
\Psi_\delta
=
\psi_{\delta/2}^{[2]}
\circ
\psi_\delta^{[1]}
\circ
\psi_{\delta/2}^{[2]},
\end{equation}
where $\psi^{[1]}$ is as in \eqref{eq:ouTransition} and $\psi^{[2]}$ as in \eqref{eq:nonlinearFlowJR}. The refined numerical transition over one observation interval of length $\Delta$ is then obtained by composing $n_{\mathrm{sub}}$ such steps:
\begin{equation}
\label{eq}
X_n^{\mathrm{OU}}
=
\underbrace{
\left(\Psi_\delta\circ\cdots\circ\Psi_\delta\right)
}_{n_{\mathrm{sub}}\ \mathrm{times}}
(X_{n-1}^{\mathrm{OU}}), \qquad n=1,\ldots,N,
\end{equation}
with independent stochastic increments in the successive substeps. The time-step refinement procedure applies analogously to the Wiener Strang splitting method \eqref{eq:wienerStrang}.

\begin{figure}[t]
    \centering
    \begin{tikzpicture}[
        >=Latex,
        node distance=0.7cm and 0.9cm,
        state/.style={
            circle,
            draw,
            minimum size=0.7cm,
            inner sep=0pt,
            font=\small
        },
        obs/.style={
            circle,
            draw,
            fill=gray!15,
            minimum size=0.7cm,
            inner sep=0pt,
            font=\small
        },
        pred/.style={
            circle,
            draw,
            dashed,
            minimum size=0.7cm,
            inner sep=0pt,
            font=\small
        },
        arrow/.style={
            ->,
            thick
        },
        subarrow/.style={
            ->,
            thick
        },
        label/.style={
            font=\small
        },
        panel/.style={
            font=\bfseries
        }
    ]

\node[panel] at (0,2.2) {(a) One-step prediction};

\node[state] (a0) at (-1.7,0.8) {$n-1$};
\node[obs]   (a1) at (0,0.8) {\footnotesize$n$};

\draw[arrow]
    (a0) -- node[above,label] {$\Delta$}
    (a1);

\node[label, below=0.75cm of a0] {filtered};
\node[label, below=0.75cm of a1] {observed};

\node[panel] at (5.0,2.2) {(b) Internal substepping};

\node[state] (b0) at (3.1,0.8) {$n-1$};
\node[state] (b1) at (4.05,0.8) {};
\node[state] (b2) at (5.0,0.8) {};
\node[state] (b3) at (5.95,0.8) {};
\node[obs]   (b4) at (6.9,0.8) {\footnotesize$n$};

\draw[subarrow]
    (b0) -- node[above,label] {$\delta$} (b1);

\draw[subarrow]
    (b1) -- node[above,label] {$\delta$} (b2);

\draw[subarrow]
    (b2) -- node[above,label] {$\delta$} (b3);

\draw[subarrow]
    (b3) -- node[above,label] {$\delta$} (b4);

\node[label, below=0.75cm of b0] {filtered};
\node[label, below=0.75cm of b4] {observed};

\draw[
    decorate,
    decoration={
        brace,
        amplitude=5pt,
        mirror
    }
]
    (3.1,0.25) -- (6.9,0.25)
    node[midway,below=0pt,label] {$n_{\mathrm{sub}}$ internal steps,
    $\delta=\Delta/n_{\mathrm{sub}}$};

\end{tikzpicture}

\caption{
Illustration of the prediction strategies used in the filtering procedure.
(a) In the standard one-step EKF, the filtered state at time $n-1$ is propagated
over one observation interval~$\Delta$ before incorporating the next given observation $y_{n}$.
(b) Internal substepping keeps the same observation interval but evaluates the
transition using $n_{\mathrm{sub}}$ finer steps of size
$\delta=\Delta/n_{\mathrm{sub}}$.}
\label{fig:predictionStrategies}
\end{figure}
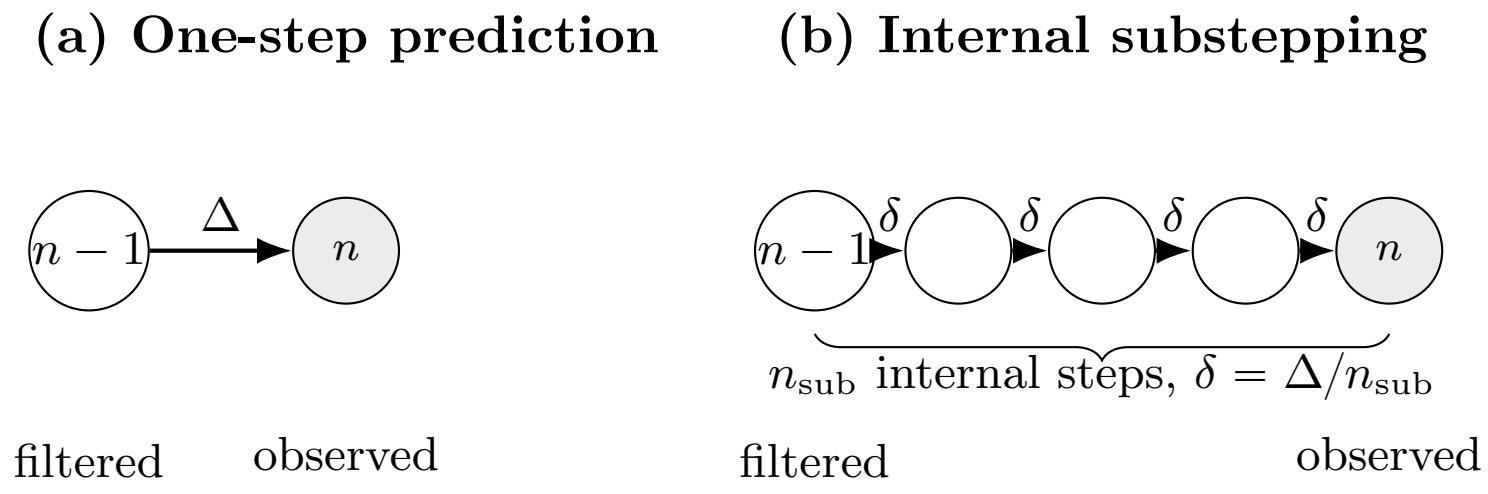

In the EKF, the predicted state mean and covariance are propagated through each internal step using the corresponding transition Jacobians and process noise covariance matrices. No additional observations are required, since the substeps are used only for numerical propagation between observation times. The EKF update is performed only when an observation is available. Figure \ref{fig:predictionStrategies} illustrates the standard one-step prediction and internal substepping procedures.

Note that the same approach can also be used to accommodate missing observations or irregular observation intervals, provided that the observation intervals can be represented by integer multiples of a sufficiently small common time step.

\section{Inference of the JR-NMM parameters from simulated~data}
\label{sec:simulation}

We assessed the performance of the proposed EKF-based inference procedures through a series of simulation experiments. In Section \ref{sec:sim_comparison}, we first compare the three EKF approaches based on different numerical discretization methods: EM-EKF, WIE-EKF, and OU-EKF. Based on this comparison, we then present the parameter estimation results for the best-performing approach, OU-EKF, in greater detail in Section \ref{sec:OU_EKF_results}. Next, we investigate the robustness of OU-EKF with respect to the observation step size, the length of the observed time series, and the measurement noise variance, including its estimation as an additional model parameter. We also examine the substepping method outlined in Section \ref{sec:substepping}. We present these investigations in Section \ref{sec:sim_robustness}. Finally, we compare the proposed OU-EKF procedure with the state-of-the-art ABC approach introduced in \cite{ireneSpringer} in Section \ref{sec:sim_abc}.

Unless stated otherwise, all simulations in this section use the parameter values in Table~\ref{tab:typicalValues}, with the stochastic JR-NMM parameters set to $\mu=220$, $\sigma=2000$, and $\tau=1$. Furthermore, we use $\Delta=2\cdot 10^{-3}$, $T=20$, and $R=0.75$. Each simulation setting is repeated $N_{rep}=1000$ times. Throughout this section, the estimated parameter vector is $\theta$ as defined in \eqref{eq:theta}, except where explicitly indicated.

To ensure that the data-generating process is impartial and provides a sufficiently accurate approximation to the underlying stochastic dynamics, we generated the data using the Euler--Maruyama scheme with step size $\Delta/100$. We then retained every $100$th simulated point, yielding observations on the coarser grid with step size $\Delta$. 

To assess the performance of the algorithms, we use the root mean squared error (RMSE) of the parameter estimates $\hat\Theta=(\hat\theta^{(1)},\ldots,\hat\theta^{(N_{rep})})\in\mathbb{R}^{3\times N_{rep}}$ relative to the true parameter vector $\theta\in\mathbb{R}^3$, where $\hat\theta^{(i)}$ denotes the estimate from simulation repetition $i$. Let $\hat\Theta_j=(\hat\theta_j^{(1)},\ldots,\hat\theta_j^{(N_{rep})})$ denote the estimates of the $j$-th parameter across all repetitions. The componentwise RMSE is then defined as
\begin{equation}\label{eq:rmse}
\operatorname{RMSE}(\hat\Theta_j;\theta_j)
= \sqrt{\frac{1}{N_{rep}}\sum_{i=1}^{N_{rep}}
\left(\hat\theta_j^{(i)}-\theta_j\right)^2},
\qquad j=1,2,3.
\end{equation}
Where appropriate, we complement these RMSEs with the 5th--95th, 10th--90th, 20th--80th, 30th--70th, and 40th--60th percentile intervals of the absolute estimation errors $|\hat\theta_j^{(i)}-\theta_j|$, $i=1,\ldots,N_{rep}$, computed separately for each parameter $j=1,2,3$, to illustrate their variability across simulation repetitions.

Moreover, we construct approximate confidence intervals for the estimated parameters using the sandwich method described in Section \ref{sec:uncertainty}. We report their empirical coverage, defined as the proportion of simulation repetitions in which the interval contains the true parameter value. Specifically, for each repetition, the coverage indicator is equal to $1$ if the true value is contained in the interval and $0$ otherwise. To quantify the uncertainty in the estimated coverage, we construct a Clopper-Pearson confidence interval \citep{clopperPearson} based on $N_{rep}=1000$ simulation~repetitions.

\subsection{Comparison of EM-EKF, WIE-EKF and OU-EKF}
\label{sec:sim_comparison}

We first compare the inference procedures obtained by combining the EKF with the OU, WIE, and EM discretizations. We compare different values of the measurement noise variance $R$ to see how the three approaches perform as the observation noise level changes.

\begin{figure}[htbp]
\centering
\includegraphics[width=\linewidth]{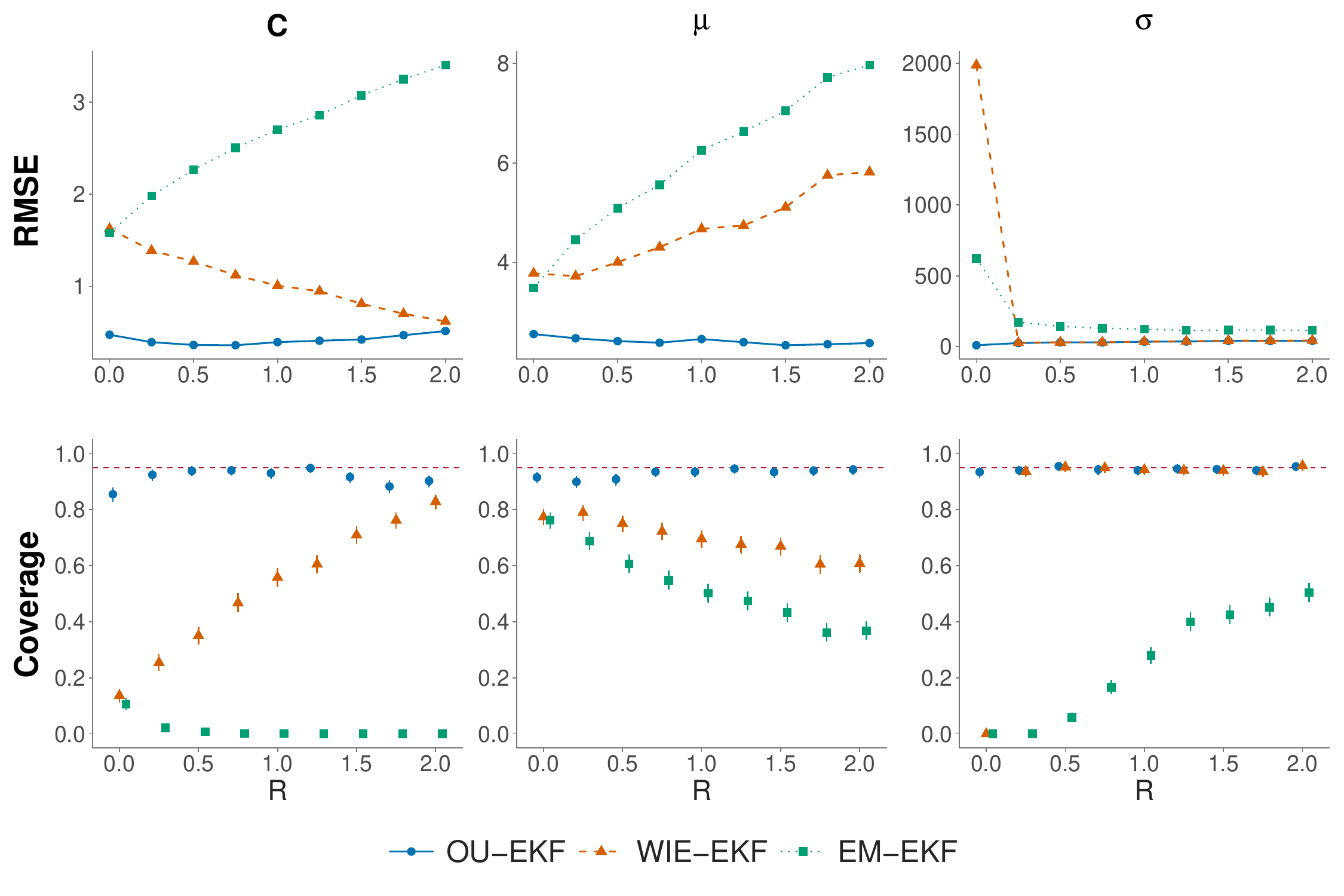}
\caption{RMSEs of the parameter estimates (top row) and empirical coverages with Clopper--Pearson confidence intervals (bottom row) for $C$, $\mu$, and $\sigma$ across different values of the measurement noise variance $R$.}
\label{comp3Fig}
\end{figure}

Figure~\ref{comp3Fig} shows the RMSEs of the resulting point estimates together with the empirical coverages of the confidence intervals, as functions of $R$. Overall, both splitting-based approaches outperform the EM-EKF, and the OU-EKF clearly outperforms the WIE-EKF. The OU-EKF yields substantially lower RMSE values and empirical coverage probabilities consistently close to the nominal level. Its performance also remains relatively stable across the considered values of $R$. The WIE-EKF generally achieves lower RMSE values and better coverage than the EM-EKF, which exhibits pronounced undercoverage. The WIE-EKF and EM-EKF show particularly large estimation errors for $\sigma$ when $R=0$ (see the top-right panel). These results indicate that, for the stochastic JR-NMM model, the OU splitting provides a more accurate basis for EKF-based parameter inference than either the WIE splitting or the standard EM discretization.

\subsection{Inference with the proposed OU-EKF method}
\label{sec:OU_EKF_results}

\begin{figure}[htbp]
\centering
\includegraphics[width=\linewidth]{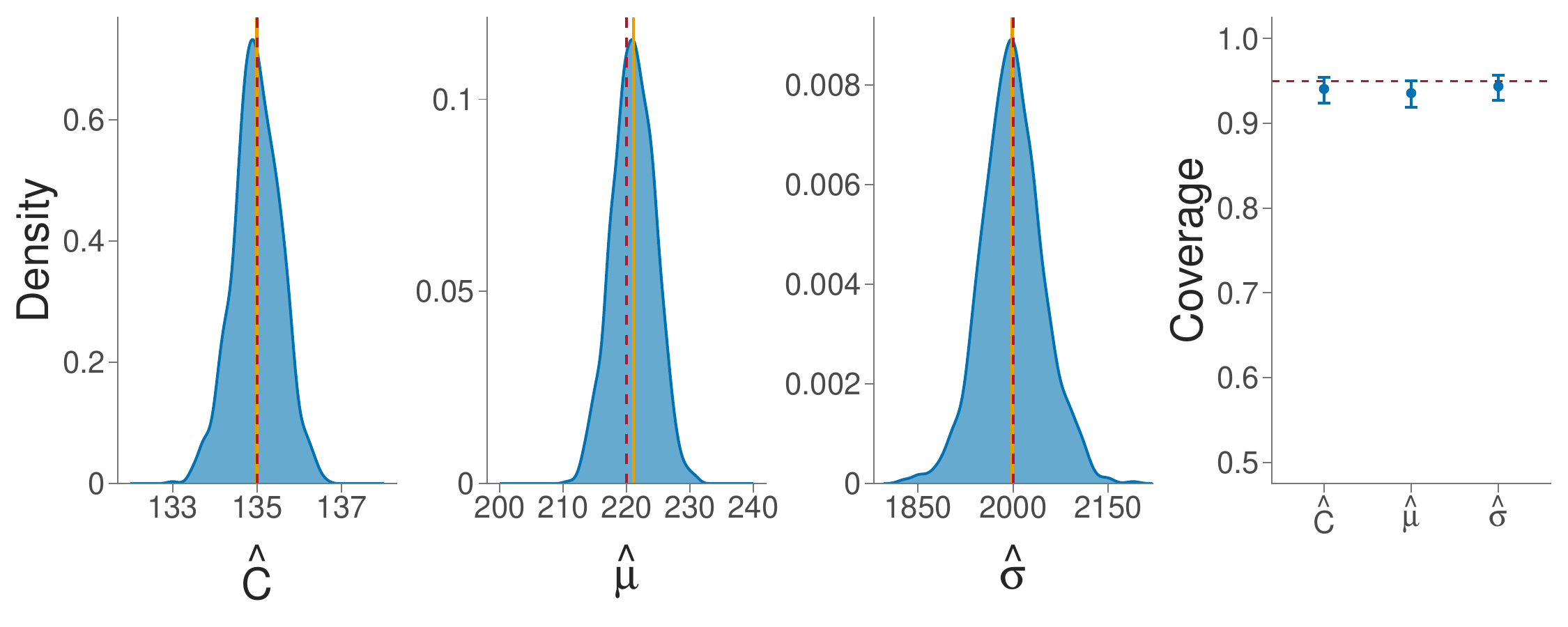}
\caption{Kernel density estimates of the OU-EKF parameter estimates for $C$, $\mu$, and~$\sigma$, with vertical red dashed lines indicating the true parameter values and vertical orange solid lines indicating the mean estimates. The right panel shows the empirical coverages with Clopper--Pearson confidence intervals.}
\label{fig:ouShow}
\end{figure}

Having established that the OU-EKF performs best among the three approaches considered, we now present its parameter estimation results in greater detail. We consider the setting with $T=20$, $\Delta=2\cdot 10^{-3}$, and $R=0.75$.

Figure~\ref{fig:ouShow} shows empirical densities of the estimates for the parameters $C$, $\mu$, and $\sigma$ across the $N_{rep}=1000$ simulation repetitions. The distributions are centered close to the respective true parameter values. This is also evident from the solid orange lines indicating the mean estimates, which are very close to the dashed red lines indicating the true parameter values. The figure also shows the empirical coverage of the corresponding confidence intervals (right panel), which is close to the nominal level of $95\%$ for all three parameters.

\begin{table}[t]
\centering
\caption{True parameter values, mean estimates, RMSEs, and empirical coverages of the confidence intervals for the OU-EKF.}
\label{tab:ouShow}
\vspace{-0.3cm}
\begin{tabular}{rllll}
  \hline
 & Parameter & C & $\mu$ & $\sigma$ \\ 
 &True Value & 135&220&2000\\
  \hline
& Mean Estimate & 134.99&221.07&1998.65\\
 & RMSE & 0.55 & 3.4 & 48.5 \\ 
   & Coverage & 94.1\% & 93.6\% & 94.4\% \\ 
   \hline
\end{tabular}
\end{table}

Table~\ref{tab:ouShow} summarizes the results numerically. The mean estimates of $134.99$, $221.07$, and $1998.65$ are close to the respective true parameter values of $135$, $220$, and $2000$ for $C$, $\mu$, and $\sigma$, respectively. The empirical coverage probabilities range from $93.6\%$ to $94.4\%$, indicating coverage close to the nominal level.

The average computation time per simulation repetition for the OU-EKF in this setting is 20.16 seconds, of which 0.21 seconds are used to simulate the data, while the inference, including the construction of confidence intervals, takes 19.94 seconds. Regarding the code used to obtain these results, we refer to the Software statement at the end of the manuscript.

\subsection{Robustness analysis}
\label{sec:sim_robustness}

Having presented the parameter estimation results for the OU-EKF in the preceding subsection, we now investigate its robustness under different simulation settings. Specifically, we examine the effects of the observation step size $\Delta$, the observation horizon $T$, and the measurement noise variance $R$, including its estimation as an additional parameter. We also investigate the effect of numerical substepping, as outlined in Section~\ref{sec:substepping}.

\begin{figure}[htbp]
\centering
\includegraphics[width=0.9\linewidth]{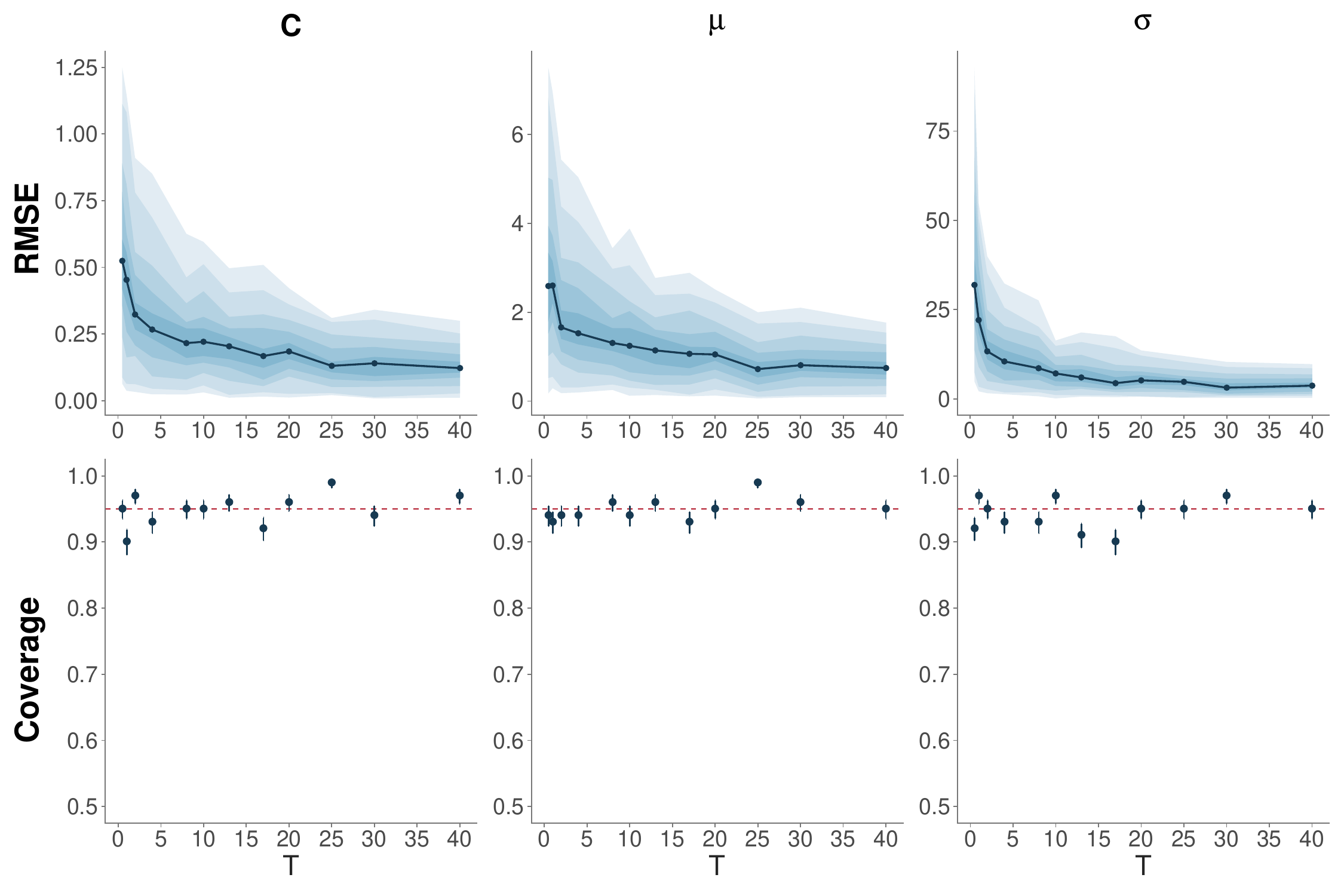}
\caption{RMSEs of the parameter estimates (top row) and empirical coverages with Clopper--Pearson confidence intervals (bottom row) for $C$, $\mu$, and $\sigma$ across different observation horizons~$T$. The shaded bands in the top row represent the percentile intervals of the absolute estimation~errors.}
\label{fig:varDelt}
\end{figure}

\paragraph{Varying the length of the time series.}
We start by varying the total observation time $T$ while keeping the remaining simulation settings fixed. Figure~\ref{fig:varDelt} shows the resulting RMSE and empirical coverage. As expected, the RMSE decreases with increasing $T$ for all three parameters. The decrease is particularly pronounced for shorter time series and becomes more gradual for larger values of $T$. The shaded bands, representing the percentile intervals of the absolute estimation errors, also become narrower as $T$ increases, indicating decreasing variability in these errors across simulation repetitions.

Moreover, empirical coverage remains close to the nominal level of $95\%$ throughout the considered range, with no systematic dependence on $T$. Overall, these results indicate that the OU-EKF provides reliable confidence interval coverage even for relatively short time series, while parameter estimation accuracy improves as more observations become available.

\paragraph{Varying the step size.}
The observation step size $\Delta$ plays a particularly important role, as it affects both the accuracy of the numerical approximation and the computational cost of the inference procedure. Larger values of $\Delta$ reduce computational cost, but may also lead to less accurate approximations. We therefore investigate how large $\Delta$ can be chosen while still obtaining reliable parameter estimates and uncertainty quantification. Since increasing $\Delta$ also reduces the number of observations for a fixed observation horizon, we consider longer time series with $T=100$ seconds to ensure sufficient data for all step sizes.

\begin{figure}[htbp]
\centering
\includegraphics[width=0.9\linewidth]{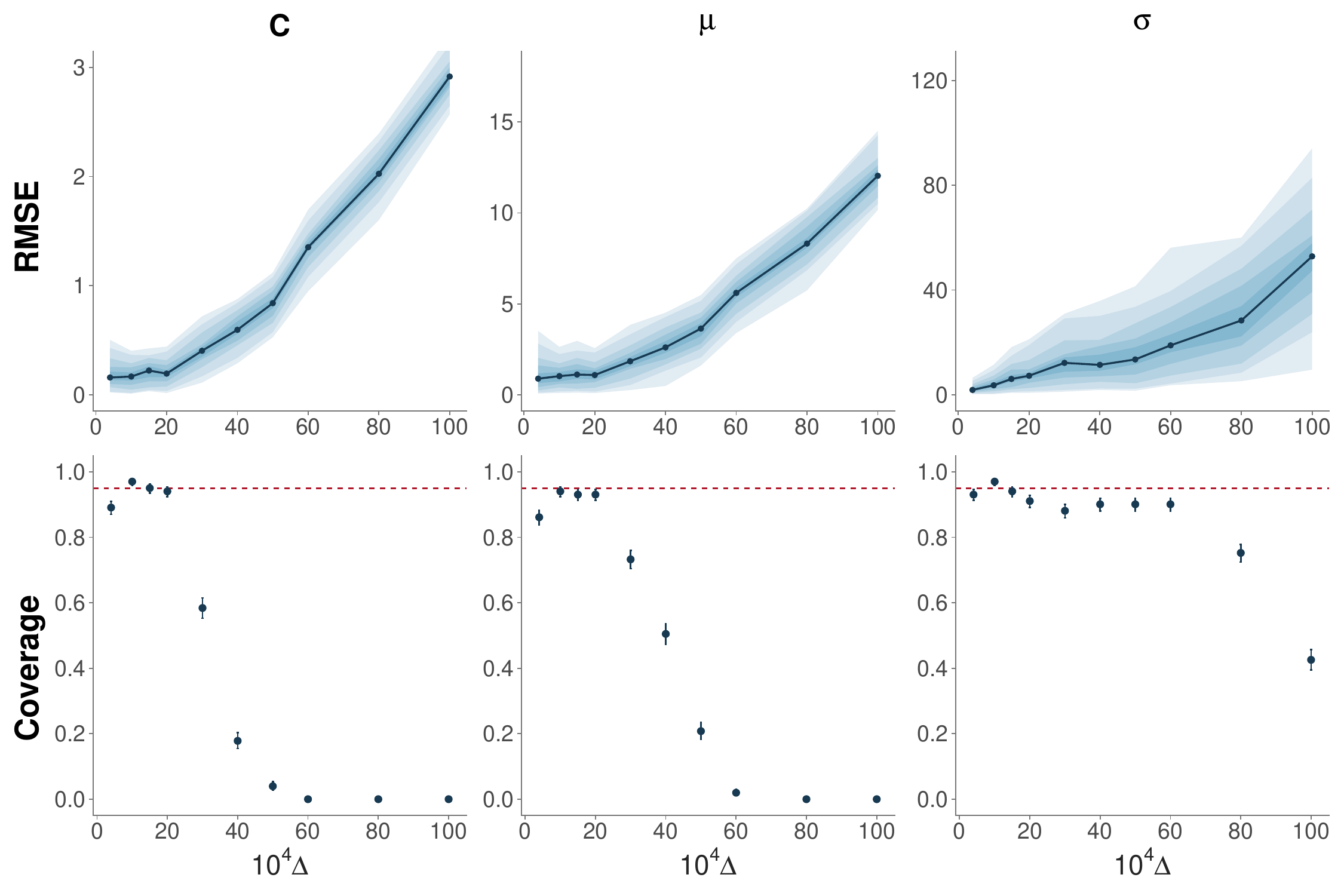}
\caption{RMSEs of the parameter estimates (top row) and empirical coverages with Clopper--Pearson confidence intervals (bottom row) for $C$, $\mu$, and $\sigma$ across different observation step sizes~$\Delta$. The shaded bands in the top row represent the percentile intervals of the absolute estimation~errors.
}
\label{fig:varC}
\end{figure}

\begin{figure}[t]
\centering
\includegraphics[width=0.9\linewidth]{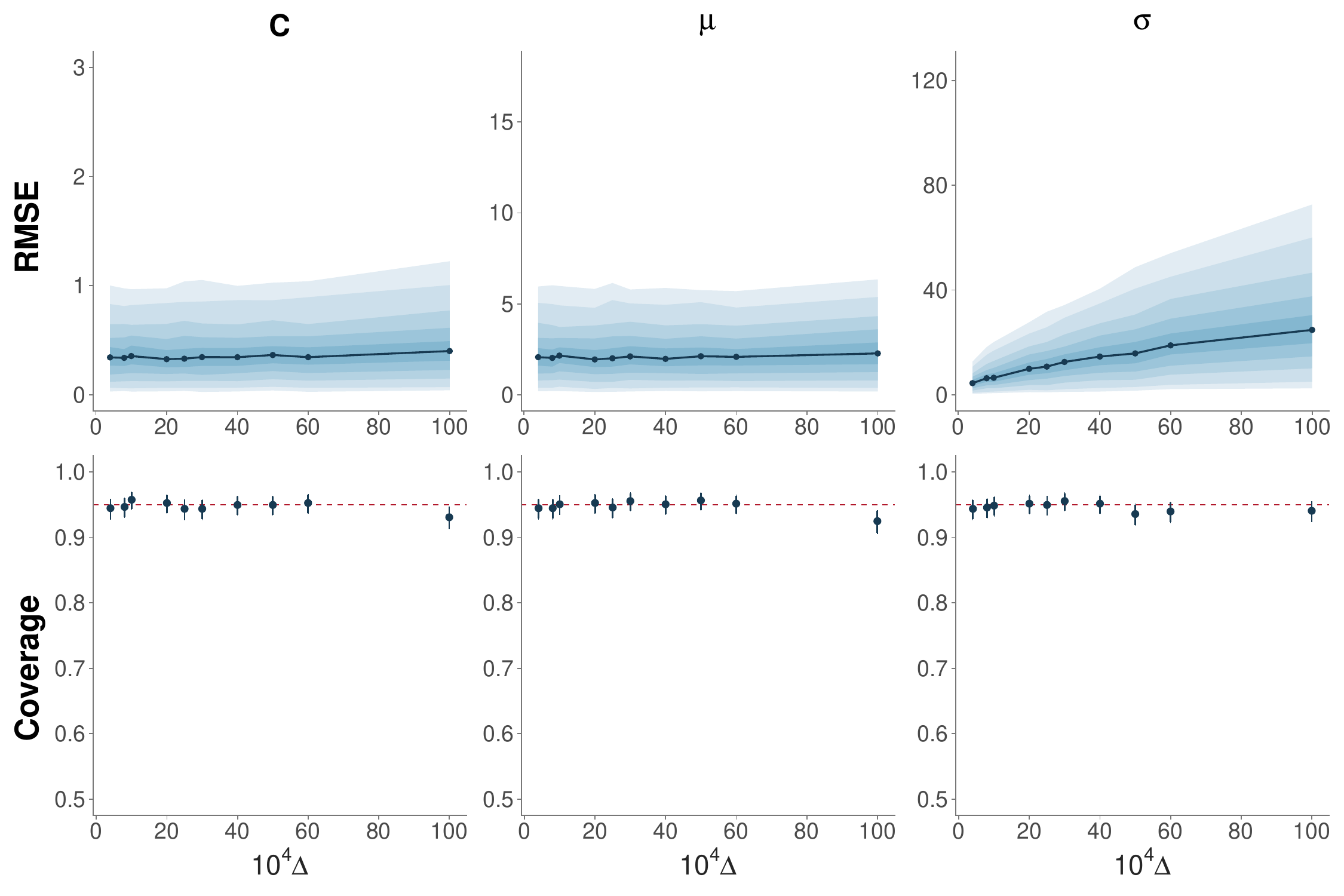}
\caption{RMSEs of the parameter estimates (top row) and empirical coverages with Clopper--Pearson confidence intervals (bottom row) for $C$, $\mu$, and $\sigma$ across different observation step sizes~$\Delta$, with substepping to an effective step size of $10^{-4}$. The shaded bands in the top row represent the percentile intervals of the absolute estimation errors.
}
\label{fig:varSubstep}
\end{figure}

Figure~\ref{fig:varC} shows the resulting RMSE and empirical coverage for different values of $\Delta$. As expected, the RMSE increases systematically with $\Delta$ for all three parameters. This behavior is consistent with the local linearization of the numerical transition map used by the EKF, as larger observation step sizes require the approximation to capture the nonlinear dynamics over longer time intervals.

The empirical coverage remains close to the nominal level of $95\%$ for observation step sizes up to $\Delta=2\cdot 10^{-3}$, but deteriorates for larger values. This decrease is particularly pronounced for $C$ and $\mu$, whereas the coverage for $\sigma$ remains relatively stable even for larger step sizes. Overall, these results suggest that the OU-EKF provides accurate parameter estimates and reliable uncertainty quantification for sufficiently small observation step sizes. At the same time, both estimation accuracy and confidence interval coverage deteriorate as $\Delta$ becomes too large.

\paragraph{Varying the step size with internal substepping.} 

We next investigate whether the substepping approach outlined in Section~\ref{sec:substepping} can improve the performance of the OU-EKF for larger observation step sizes. We again vary the observation step size $\Delta$, but now use internal substepping to an effective step size of $10^{-4}$. Figure~\ref{fig:varSubstep} shows the resulting RMSE and empirical coverage for different $\Delta$ values.

Compared with the results without substepping (cf. Figure \ref{fig:varC}), the improvement in estimation accuracy is clear. The RMSE remains nearly constant across the considered step sizes for $C$ and $\mu$. For $\sigma$, the RMSE increases with $\Delta$, but considerably more slowly than without substepping. Moreover, empirical coverage remains close to the nominal $95\%$ level throughout the considered range of observation step sizes.

Overall, these results indicate that substepping can substantially improve the OU-EKF's robustness to larger observation step sizes while maintaining reliable uncertainty quantification.

\paragraph{Varying and estimating the measurement noise.}

\begin{figure}[htbp]
\centering
\includegraphics[width=0.9\linewidth]{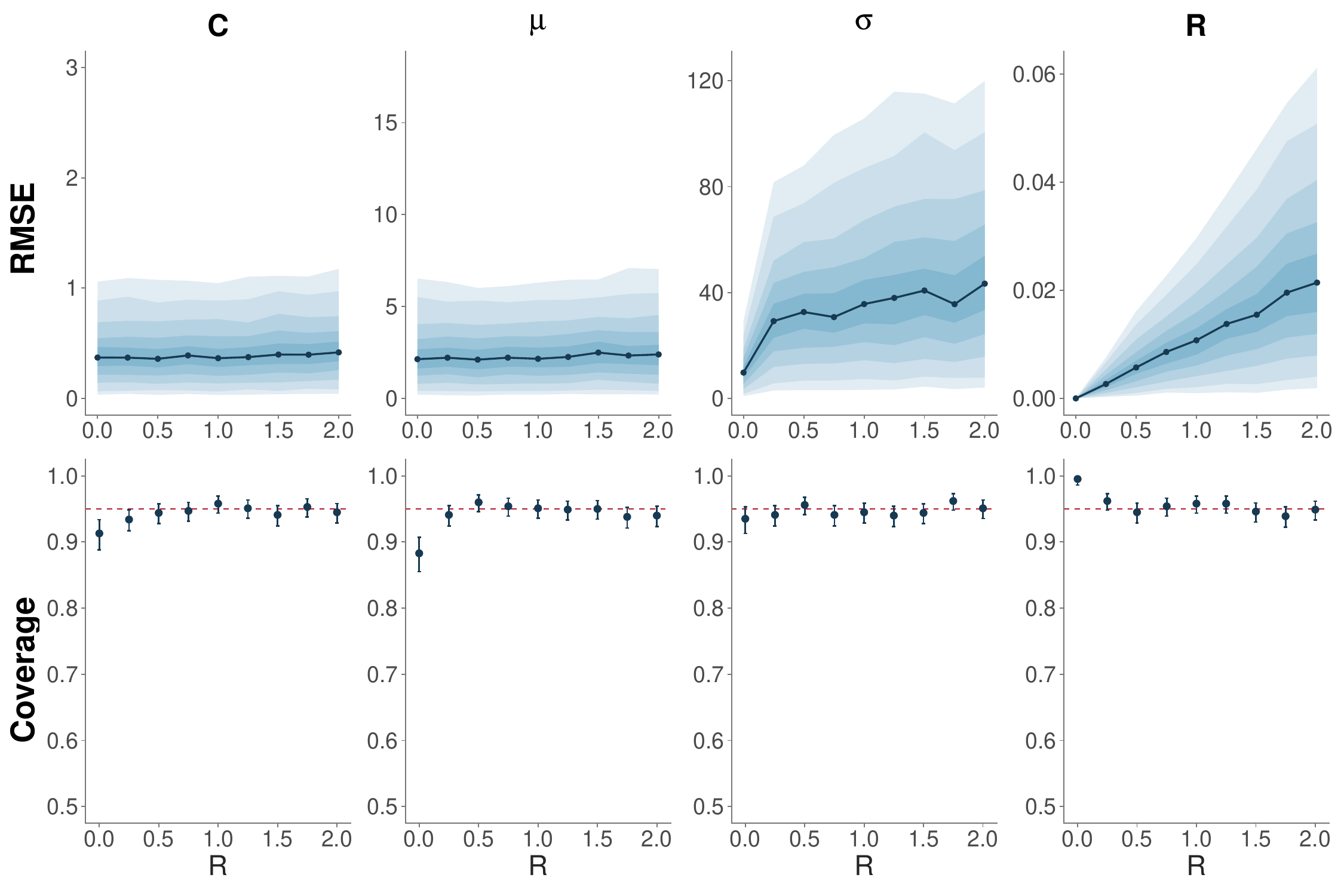}
\caption{RMSEs (top row) and empirical coverages with Clopper--Pearson confidence intervals (bottom row) for $C$, $\mu$, $\sigma$, and $R$ across different true values of the measurement noise variance~$R$, which is estimated as an additional parameter. The shaded bands in the top row represent the percentile intervals of the absolute estimation errors.}
\label{fig:varR}
\end{figure}

\begin{figure}[htbp]
\centering
\includegraphics[width=\linewidth]{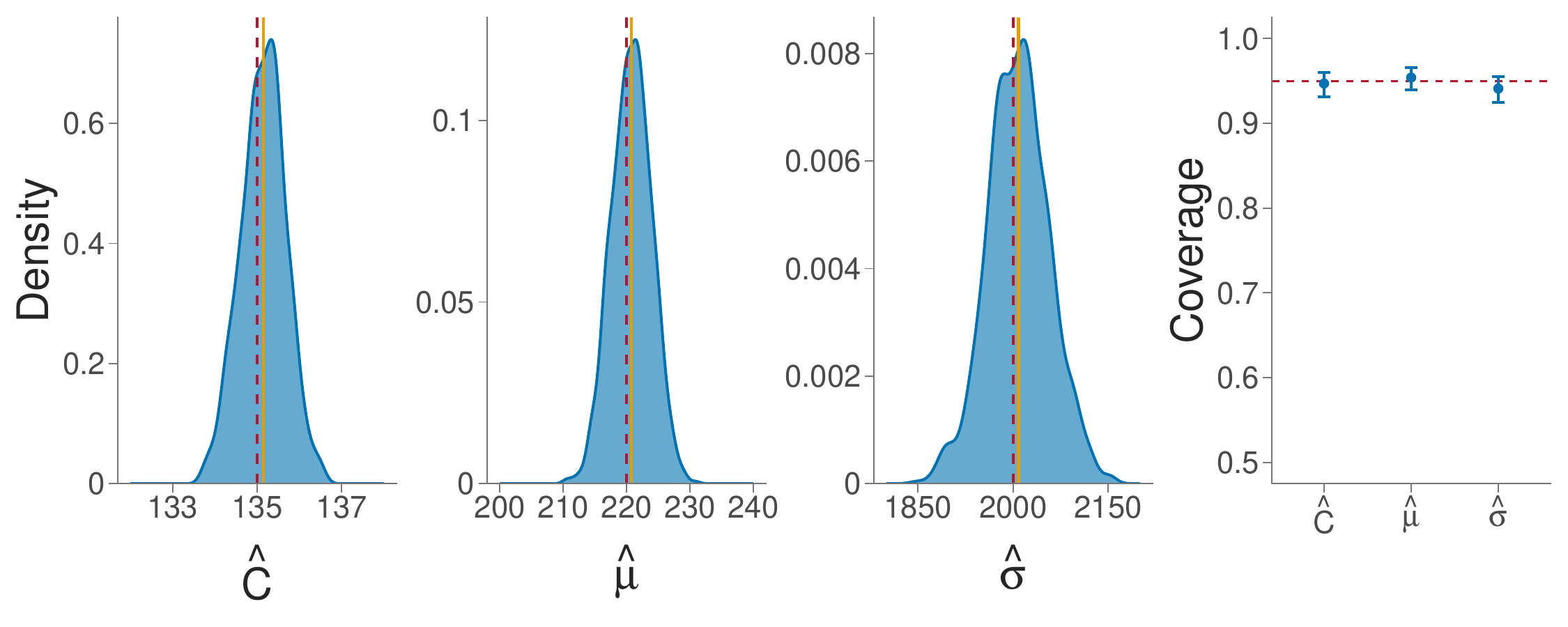}
\caption{Kernel density estimates of the OU-EKF parameter estimates for $C$, $\mu$, and $\sigma$, with $R$ additionally estimated (true value $R=0.75$). The vertical red dashed lines and solid orange lines indicate the true parameter values and mean estimates, respectively. The right panel shows empirical coverage with Clopper--Pearson confidence intervals.}
\label{fig:ouShow_R}
\end{figure}

We next investigate the performance of the OU-EKF when we estimate the measurement noise variance $R$ as an additional parameter. Figure~\ref{fig:varR} shows the RMSE and empirical coverage for $C$, $\mu$, $\sigma$, and $R$ across different true values of the measurement noise variance. The RMSE for $R$ increases with its true value, indicating that its estimation becomes less accurate at higher noise levels. For $R=0$, the estimates are particularly accurate, with very narrow confidence intervals. The estimation accuracy of $C$ and $\mu$ remains relatively stable across the considered values of $R$, whereas the estimation of $\sigma$ becomes less accurate for larger values of $R$. This is to be expected, since increasing measurement noise makes it more difficult to infer the variability arising from the stochastic model dynamics. Nevertheless, the empirical coverage remains close to the nominal $95\%$ level for all  parameters.

To examine the effect of estimating $R$ on the remaining parameters in greater detail, we focus on the setting with $R=0.75$. Figure~\ref{fig:ouShow_R} shows the empirical densities of the estimates for $C$, $\mu$, and $\sigma$, which are similar to those obtained when $R$ was treated as known in Section~\ref{sec:OU_EKF_results} (cf. Figure~\ref{fig:ouShow}). Again, the distributions are centered close to the respective true parameter values, with the solid orange lines indicating the mean estimates lying close to the dashed red lines indicating the true values. The numerical results in Table~\ref{tab:ouShow_R} support this observation. The mean estimates of $135.15$, $220.77$, and $2008.66$ are close to the respective true values of $135$, $220$, and $2000$ for $C$, $\mu$, and $\sigma$, respectively. The RMSE values of $0.55$, $3.21$, and $48.8$, as well as the empirical coverage probabilities of $94.7\%$, $95.4\%$, and $94.1\%$, are also comparable to those reported in Table~\ref{tab:ouShow}.

The similar results with regards to the previous experiment suggest that estimating $R$ as an additional parameter does not substantially affect the inference performance of OU-EKF for $C$, $\mu$, and $\sigma$ under the considered settings.

\begin{table}[t]
\centering
\caption{True parameter values, mean estimates, RMSEs, and empirical coverages of the confidence intervals for the OU-EKF.}
\label{tab:ouShow_R}
\vspace{-0.3cm}
\begin{tabular}{rllll}
  \hline
 & Parameter & C & $\mu$ & $\sigma$ \\ 
 &True Value & 135&220&2000\\
  \hline
& Mean Estimate & 135.15&220.77&2008.66\\
 & RMSE & 0.55 & 3.21 & 48.8 \\ 
   & Coverage & 94.7\% & 95.4\% & 94.1\% \\ 
   \hline
\end{tabular}
\end{table}

\subsection{Comparison to the state-of-the-art ABC method}
\label{sec:sim_abc}

Finally, we compare the proposed OU-EKF inference procedure with the ABC method introduced in \citet{ireneSpringer}. While their approach uses standard accept--reject ABC, we employ a more computationally efficient sequential Monte Carlo (SMC) ABC algorithm. Details of the SMC-ABC implementation are provided in the accompanying code (see the Software statement at the end of the manuscript). As in the preceding experiments, we estimate the parameter vector~$\theta$ defined in \eqref{eq:theta}. We set $R=0$, since \citet{ireneSpringer} did not consider measurement noise. We base the comparison on $1000$ simulation repetitions for the OU-EKF and, due to the higher computational cost, $100$ for SMC-ABC.

\begin{figure}[t]
\centering\includegraphics[width=1.0\linewidth]{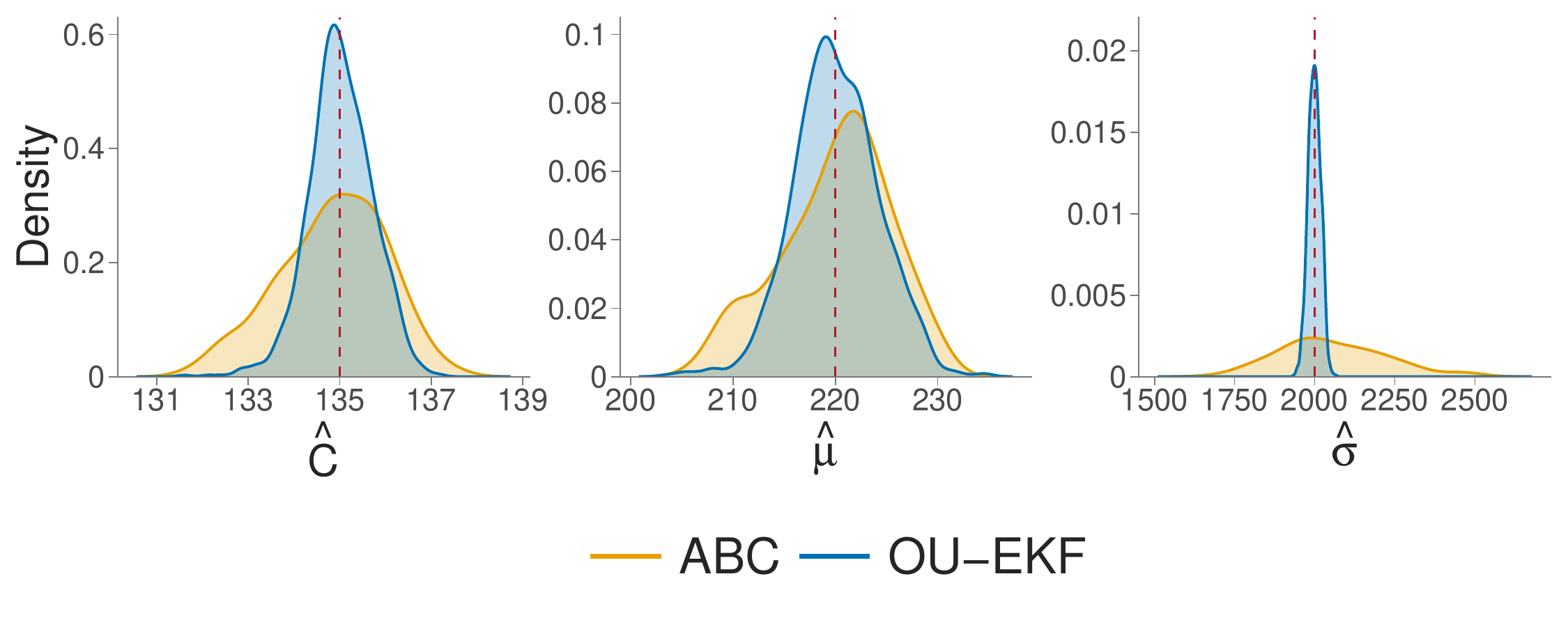}
\caption{Kernel density estimates of the parameter estimates for $C$, $\mu$, and $\sigma$ obtained using the ABC and OU-EKF inference procedures, with vertical dashed lines indicating the true parameter values. For the ABC approach, the considered point estimates correspond to the weighted posterior means.}
\label{abcEkfFig}
\end{figure}

Figure~\ref{abcEkfFig} shows kernel density estimates of the parameter estimates across simulation repetitions, along with the corresponding true parameter values. For SMC-ABC, the point estimates correspond to the weighted posterior means. For all three parameters, the OU-EKF estimates show less dispersion than the SMC-ABC estimates. Moreover, they are better centered around the respective true parameter values. These differences are particularly pronounced for $\sigma$. The RMSE values reported in Table~\ref{rmseAbcEkfCompTable} confirm this finding, with the OU-EKF achieving significantly lower RMSEs for all three parameters.

Beyond its higher estimation accuracy, the OU-EKF offers a considerable computational advantage. Even with the SMC extension, the ABC procedure requires substantially more computation time than the OU-EKF at the same observation step size. Furthermore, the robustness experiments in Section~\ref{sec:sim_robustness} demonstrate that the OU-EKF can provide reliable inference even for relatively short time series. By comparison, the ABC approach relies on summary statistics that may be less stable when computed from short time series \citep{ireneSpringer}. These results suggest that the OU-EKF offers better estimation accuracy and lower computational cost, while also applying to shorter observed time series.

\begin{table}[t]
\centering\caption{RMSEs for ABC and OU-EKF.}
\vspace{-0.3cm}
\begin{tabular}{rrrr}
\hline
RMSE & C & $\mu$ & $\sigma$ \\
\hline
ABC & 1.17 & 5.35 & 174.05 \\
OU-EKF & 0.70 & 4.17 & 19.52 \\
\hline
\end{tabular}\label{rmseAbcEkfCompTable}
\end{table}

\section{Inference of the JR-NMM parameters from real EEG~data}
\label{sec:application}

In this section, we apply the proposed {OU-EKF} inference method to a real EEG time series dataset to estimate the parameter vector $\theta$ defined in \eqref{eq:theta} of the stochastic JR-NMM \eqref{eq:jrSDE}. 

We consider real EEG data from the publicly available collection described by \citet{Andrzejak2001} and accessible \href{https://repositori.upf.edu/items/77d77b11-1254-4fc2-9065-e37e6ff6e217}{online} (see also the Data Availability statement at the end of the manuscript). Specifically, we focus on recording \texttt{O019}, which \citet{ireneSpringer} previously investigated in the context of ABC inference (see Figure $10$ therein). This signal exhibits alpha-rhythmic behavior, associated with connectivity parameter values of $C \approx 135$ in the JR-NMM (cf. Section~\ref{sec:jr}). The recording was sampled at $173.61$ Hz,
corresponding to an observation interval of approximately $\Delta=0.00576$. Since this interval is relatively large for the numerical approximation of the stochastic JR-NMM (cf. \cite{haraldPaper}), we
therefore employ the time-step refinement procedure described in Section~\ref{sec:substepping}, using $n_{\mathrm{sub}}=10$ internal propagation steps per
observation interval.

Unlike in the simulation study of Section~\ref{sec:simulation}, the stochastic JR-NMM provides only an approximate representation of the observed EEG signal. Specifically, the model describes the interactions of a small number of neural groups and cannot capture all features of measured EEG activity \citep{jansenRit,computationalEEG}. Furthermore, the model output is not directly equivalent to scalp potentials, which depend on the spatial distribution of neural sources and volume conduction through the head \citep{EEGSourceReview}. These limitations introduce potential sources of model discrepancy that may affect EKF-based parameter inference.

To investigate the sensitivity of the inferred parameters to different ways of accounting for such model discrepancies, we consider three modifications of the standard {OU-EKF} approach:
\begin{itemize}
    \item \textbf{Increased observation uncertainty}: We consider a range of strictly positive values of the observation-noise variance $R$ in \eqref{eq:jrObservationEKF}, including larger values to allow for greater discrepancies between the measured EEG signal and the EEG-like output of the stochastic JR-NMM, and examine the sensitivity of the parameter estimates to these prescribed values of $R$.
    \item \textbf{ACF-penalized approximate likelihood}: We augment the approximate innovation likelihood \eqref{eq:quasiLikelihood} with a penalty based on the autocorrelation function (ACF), encouraging agreement between the temporal dependence structure of the observed and model-generated signals.
    \item \textbf{Estimation of the process-noise scale}: We include the additional process-noise parameter $\tau$ of the stochastic JR-NMM \eqref{eq:jrSDE} in the parameter vector $\theta$ to be estimated, allowing greater flexibility in the model's diffusion coefficient.
\end{itemize}

These three modifications address different aspects of model discrepancy and are introduced in Sections \ref{sec:application_observation}–\ref{sec:noise_scale}, respectively, together with their corresponding parameter estimates. Section~\ref{sec:comparison_EEG} then compares these estimates with those obtained using the standard {OU-EKF} approach with $R=0$ and assesses the \textit{goodness of fit} of the different methods.

\subsection{Allowing for increased observation uncertainty}
\label{sec:application_observation}

When applying the {OU-EKF} method to real EEG data, the observation-noise variance $R$ in~\eqref{eq:jrObservationEKF} becomes particularly relevant because of measurement uncertainty and model discrepancy.
We investigate how the inferential results change when $R$ is fixed at different strictly positive values.

Rather than estimating $R$ jointly with the model parameters $\theta$, we prescribe its value to ensure that the filtering procedure allows for a certain degree of observation uncertainty. Indeed, when $R$ is included as an additional parameter to be estimated, the approximate likelihood favors values close to zero in this application. By fixing $R>0$, we can instead examine the sensitivity of the parameter estimates to a progressively less restrictive observation model. In particular, increasing $R$ increases the innovation covariance $S_n$ (cf. \eqref{eq:innovationCov} and \eqref{eq:jrInnovationCov}) and thereby reduces the Kalman gain $K_n$ (cf. \eqref{eq:kalmanGain}). Consequently, the observations have less influence on the EKF state updates.

\begin{figure}[t]
\centering
\includegraphics[width=.7\linewidth]{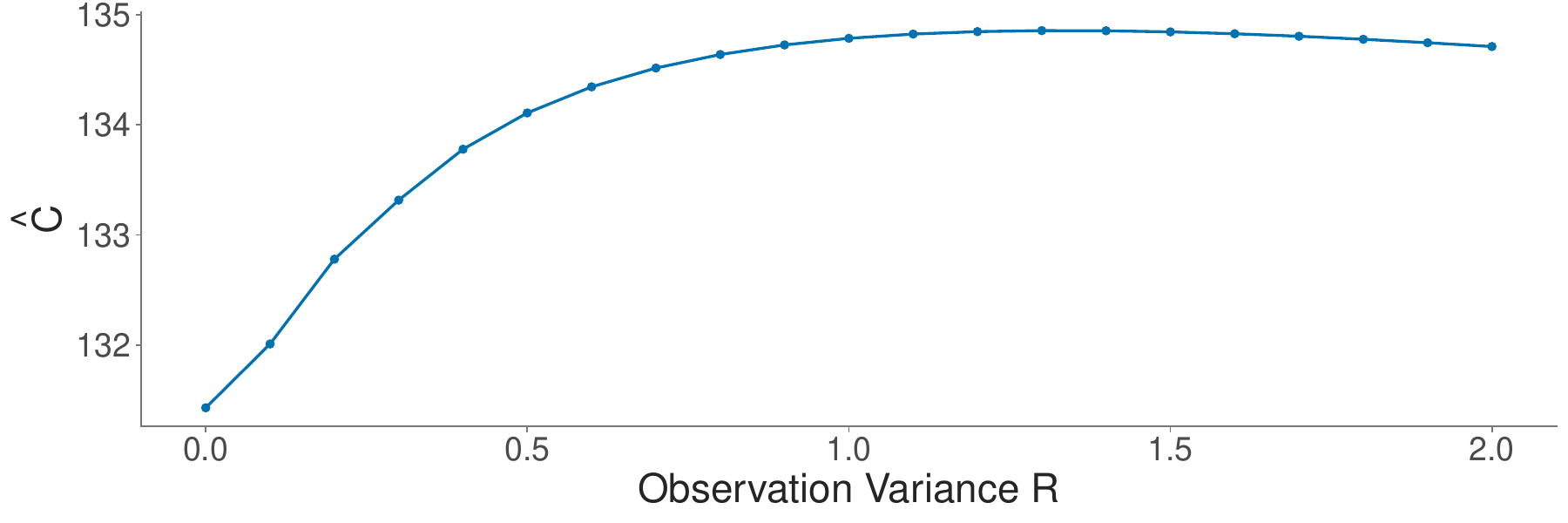}
\caption{
Estimate of $C$ obtained under different values of the
observation-noise variance $R$. 
}
\label{fig:cByVar}
\end{figure}

Figure~\ref{fig:cByVar} shows the resulting estimate of $C$ as a function of the prescribed observation-noise variance $R$. We focus on $C$ because it strongly influences JR-NMM dynamics, with values around $C=135$ associated with alpha-rhythmic behavior (cf. Section~\ref{sec:jr}). As $R$ increases, the estimated value of $C$ changes systematically and stabilizes around $C=134.5$ for larger observation-noise variances. These results suggest that the inferred connectivity parameter is sensitive to the observation-noise specification when $R$ is small, but becomes more stable as we allow greater observation uncertainty. Thus, the estimate is not specific to a single choice of~$R$, but remains approximately constant over a range of increased observation uncertainties.

Based on this sensitivity analysis, we select $R=2$ as a representative value and denote the resulting modified method by $\textbf{OU-EKF}_{\textrm{unc}}$. This method corresponds to the standard OU-EKF approach with the observation-noise variance fixed at $R=2$ and yields the following estimates for~$\theta$ defined in \eqref{eq:theta}:
$$
    \hat C=134.71,\qquad
    \hat\mu=222.21,\qquad
    \hat\sigma=2144.9.
$$

The estimated value of $C$ is close to $135$, which is associated with alpha-rhythmic behavior in the JR-NMM. Furthermore, all three parameter estimates agree well with the corresponding ABC-based estimates reported by \citet{ireneSpringer} (see Figure~11 therein). We emphasize that the prescribed value $R=2$ is not intended as an estimate of the actual measurement-noise variance, but is used to assess the effect of allowing for discrepancies between the measured EEG signal and the EEG-like output of the stochastic JR-NMM.

\subsection{ACF-penalized approximate innovation likelihood}
\label{sec:ACF_penalty}

Another possible source of model discrepancy concerns the autocorrelation structure of the observed EEG signal. The approximate innovation likelihood is based on the one-step-ahead prediction errors obtained from the EKF. It does not explicitly require the stochastic JR-NMM to reproduce the autocorrelation function (ACF) of the observed data. Consequently, parameter estimates obtained by maximizing this likelihood may yield simulated EEG-like signals whose autocorrelation structure differs from that of the measured EEG signal. 

To investigate this aspect of model discrepancy, we augment the approximate innovation likelihood with a criterion that measures the difference between the ACFs of the observed EEG signal and simulated EEG-like signals from the stochastic JR-NMM. Let
$$
    {\rho}_{\mathrm{obs}}(h),
    \qquad h=1,\ldots,L,
$$
denote the empirical ACF of the observed EEG signal at lag $h$. For a candidate parameter vector~$\theta$, let ${\rho}^{(m)}_{\theta}(h)$ denote the empirical ACF of the $m$th simulated EEG-like trajectory of the stochastic JR-NMM under $\theta$. We define
\begin{equation}
\label{eq:applicationACF}
D_{\mathrm{ACF}}(\theta)
=
\frac{1}{M}
\sum_{m=1}^{M}
\sum_{h=1}^{L}
\left[
{\rho}^{(m)}_{\theta}(h)
-
{\rho}_{\mathrm{obs}}(h)
\right]^2 .
\end{equation}
For each candidate parameter vector $\theta$, we simulate the $M$ trajectories used to evaluate this ACF discrepancy using the same set of underlying random-number sequences, so different random draws do not cause changes in the discrepancy during optimization.

The resulting estimation criterion is
\begin{equation}
\label{eq:applicationPenalized}
\ell_{\mathrm{pen}}(\theta\mid y_{1:N})
=
\ell(\theta\mid y_{1:N})
-
\lambda D_{\mathrm{ACF}}(\theta),
\end{equation}
where $\ell(\theta\mid y_{1:N})$ is the approximate innovation log-likelihood defined in \eqref{eq:quasiLikelihood} and $\lambda>0$ controls the weight of the ACF discrepancy term. This additional term is not interpreted as a likelihood contribution, but as a penalty that encourages agreement between the autocorrelation structures of the observed EEG signal and the simulated EEG-like signals.

We denote the resulting modified inference method by $\textbf{OU-EKF}_{\textrm{ACF}}$, with the observation-noise variance fixed at $R=0$. It yields the following parameter estimates:
$$
    \hat C=134.54,\qquad
    \hat\mu=225.61,\qquad
    \hat\sigma=2077.6.
$$

The estimated value of $C$ is again close to $135$, which is associated with alpha-rhythmic behavior in the JR-NMM. Moreover, all three parameter estimates agree well with those obtained using the $\text{OU-EKF}_{\mathrm{unc}}$ method in Section~\ref{sec:application_observation}, as well as with the ABC-based estimates reported by \citet{ireneSpringer} (see Figure~11 therein).

\subsection{Estimating the process-noise scale}
\label{sec:noise_scale}

Another way to account for model discrepancy is to allow greater flexibility in the diffusion coefficient of the stochastic JR-NMM \eqref{eq:jrSDE}. In the standard OU-EKF approach, the process-noise parameter $\tau$ is fixed at $\tau=1$, while $\sigma$ is estimated as part of the parameter vector $\theta$. Recall that $\tau$ controls the noise intensity in the first and third velocity equations, whereas $\sigma$ controls the noise intensity in the second velocity equation (cf. Section~\ref{sec:jr:model}).

We now consider estimating $\tau$ jointly with the three previously considered parameters, resulting in the extended parameter vector
\begin{equation*}
\theta=(C,\mu,\sigma,\tau).
\end{equation*}
This modification allows parameter estimation to adjust the noise intensities in all three velocity components. We denote the resulting inference method by $\textbf{OU-EKF}_{\textrm{tau}}$, with the observation-noise variance again fixed at $R=0$. The corresponding parameter estimates are
$$
    \hat C=134.43,\qquad
    \hat\mu=159.27,\qquad
    \hat\sigma=1320,\qquad
    \hat\tau=44.27.
$$

The estimated value of $\tau$ is substantially larger than its previously fixed value of $1$. At the same time, the estimate of $\sigma$ decreases compared with the previous approaches, indicating that allowing additional stochastic forcing in the first and third velocity components also affects the inferred noise intensity in the second component. The estimate of $\mu$, the drift parameter of the stochastic external input, is likewise smaller. Interestingly, despite these changes, the estimated value of $C$ remains close to $135$ and agrees well with the estimates obtained using the $\text{OU-EKF}_{\textrm{unc}}$ and $\text{OU-EKF}_{\textrm{ACF}}$ methods in the previous subsections.

\subsection{Comparison of parameter estimates and goodness of fit}
\label{sec:comparison_EEG}

We now compare the parameter estimates obtained using the three modifications considered in the previous subsections with those from the standard $\text{OU-EKF}$ approach with $R=0$. In addition to the parameter estimates, we assess how well the corresponding fitted models reproduce selected statistical properties of the observed EEG signal.

For this purpose, we consider two goodness-of-fit measures. The first compares the empirical densities of consecutive differences in the observed and predicted signals. Specifically, we run the Kalman filter under the estimated parameter vectors and record the predicted observation (cf. $\bar y_n$ in \eqref{eq:ekfObservationPrediction}). Each increment (=one-step) density is then compared with the corresponding increment density of the observed EEG signal using the Kullback-Leibler divergence. We report these distances as the \textit{Increment KLD}.

The second measure assesses agreement in the autocorrelation structure. Specifically, for each estimated parameter vector, we generate $100$ simulated trajectories and compute the sum of squared differences between its empirical ACF and that of the observed EEG signal. We report the median of these 100 discrepancies as the \textit{ACF error}. For both measures, smaller values indicate better agreement with the observed EEG signal.

\begin{table}[t] \centering \caption{Parameter estimates and goodness-of-fit measures for the observed EEG signal.} \label{tab:applicationComparison} \begin{tabular}{lcccccc} \hline Method & $C$ & $\mu$ & $\sigma$ & $\tau$ & Increment KLD & ACF error \\ \hline \text{OU-EKF} &120.44 &161.37 &2551.1 &1 &0.0056 &2.92 \\ $\text{OU-EKF}_{\textrm{unc}}$ &134.71 &222.21 &2144.9 &1 &0.0054 &0.51 \\ 
 $\text{OU-EKF}_{\textrm{ACF}}$ &134.54 &225.61 &2077.6 &1 &0.0054&\text{0.35} \\ 
 $\text{OU-EKF}_{\textrm{tau}}$ &134.43 &159.27 &1320 &44.27 &0.\text{0051} &7.72 \\ \hline \end{tabular} \end{table}

Table~\ref{tab:applicationComparison} summarizes the parameter estimates and the two goodness-of-fit measures for all four approaches. The parameter estimates reveal a clear difference between the standard OU-EKF approach and the three modifications. While the standard approach yields $\hat C=120.44$, all three modifications produce estimates close to $134.5$, in agreement with the value $C\approx135$ associated with alpha-rhythmic behavior in the JR-NMM. The estimates obtained using $\text{OU-EKF}_{\textrm{unc}}$ and $\text{OU-EKF}_{\textrm{ACF}}$ are particularly close, not only for $C$ but also for $\mu$ and $\sigma$. As noted in the previous subsections, these estimates also agree well with the ABC-based results reported by \citet{ireneSpringer} (see Figure~11 therein).

The goodness-of-fit measures provide further insight into the differences between the approaches. All three modifications yield slightly smaller Increment KLDs than the standard OU-EKF fit, although the differences are small. Overall, the Increment KLDs are very small for all four approaches, which is consistent with the EKF-based inference framework, which relies on one-step-ahead predictions. More pronounced differences arise in the ACF error. Both $\text{OU-EKF}_{\textrm{unc}}$ and $\text{OU-EKF}_{\textrm{ACF}}$ substantially improve the agreement in the autocorrelation structure compared with the standard fit, with the ACF-penalized approach yielding the smallest discrepancy.  

In contrast, estimating $\tau$ leads to the largest ACF error among the four approaches, despite producing an estimate of $C$ close to those of the other modifications and a similarly small Increment KLD. Thus, the additional flexibility in the diffusion coefficient does not lead to a better fit according to both measures considered here. In particular, the results favor the two approaches that retain the fixed value $\tau=1$ while modifying either the observation model or the estimation criterion. This does not imply that estimating $\tau$ is generally disadvantageous, but it shows that including it is not sufficient to improve agreement with the observed autocorrelation structure in this application.

Overall, the results show that how you account for model discrepancy can substantially affect the inferred parameters of the stochastic JR-NMM using the OU-EKF approach. Notably, three different modifications yield closely agreeing estimates of $C$ near the value associated with alpha-rhythmic behavior. Among these, increasing the observation uncertainty and incorporating an ACF penalty also provide the strongest overall agreement with the observed EEG signal according to the measures considered. These findings illustrate the value of accounting for model discrepancy when fitting the stochastic JR-NMM to real EEG data using EKF-based parameter estimation.

\section{Conclusions and outlook}
\label{sec:discussion}

In this work, we have developed a filtering-based approach to parameter inference for the stochastic Jansen--Rit neural mass model (JR-NMM), combining the extended Kalman filter (EKF) with structure-preserving numerical splitting methods. We placed particular emphasis on the Ornstein--Uhlenbeck splitting method (OU), which combines the linear drift dynamics and the diffusion term into an exactly solvable stochastic subproblem. The resulting OU-EKF benefits from a full-rank transition covariance matrix and explicitly accounts for the nontrivial process-noise Jacobian when propagating predictive uncertainty. These features distinguish the OU-EKF from the corresponding Wiener splitting (WIE-EKF) and Euler--Maruyama (EM-EKF) formulations and illustrate how the choice of numerical discretization affects the construction of the approximate likelihood.

Our simulation study demonstrates the advantages of using structure-preserving splitting methods for parameter inference in the stochastic JR-NMM. Both splitting-based approaches outperform the standard EM-EKF, with the OU-EKF providing the best inferential results among the three methods considered. Further simulation experiments demonstrate the OU-EKF's robustness across different observation horizons, observation step sizes, and measurement noise levels. The method provides accurate parameter estimates and reliable uncertainty quantification over a broad range of observation settings. Estimation accuracy improves with increasing observation horizon, while the method achieves reliable confidence-interval coverage even for relatively short time series. Moreover, even for relatively large observation step sizes, accurate inference can be achieved through internal substepping without requiring more frequent observations. The results also indicate that reliable inference for the model parameters can be maintained when the measurement noise variance $R$ is estimated as an additional parameter. The comparison with the state-of-the-art sequential Monte Carlo approximate Bayesian computation (SMC-ABC) approach further demonstrates the proposed method's effectiveness. In the simulation settings considered, the OU-EKF achieves higher parameter estimation accuracy while requiring substantially less computational effort. 

We further applied the OU-EKF to real EEG data exhibiting alpha-rhythmic activity. In this application, we investigated several approaches to accounting for discrepancies between the observed EEG signal and the EEG-like output of the stochastic JR-NMM, including modifications of the observation uncertainty, the stochastic forcing, and the likelihood criterion. These modifications yielded closely agreeing estimates of the principal dynamical parameter despite accounting for model discrepancy in different ways. The resulting parameter estimates were physiologically plausible and consistent with established findings on alpha rhythms in the literature. This application demonstrates the practical applicability of the OU-EKF to real EEG recordings and the importance of appropriately accounting for model discrepancy in parameter inference.

Several directions for future research arise from the present work. One possibility is to investigate alternative filtering-based approaches to parameter inference for the stochastic JR-NMM. In particular, approaches based on joint state and parameter estimation, in which unknown parameters are incorporated into an augmented state vector, could provide an alternative to the approximate likelihood optimization considered here. Other nonlinear filtering methods, such as the unscented Kalman filter, ensemble Kalman filter, or particle filters, could also be investigated in combination with structure-preserving numerical splitting methods.

A particularly promising extension concerns the multi-population stochastic JR-NMM, in which several neuronal populations interact through a network of connections. In this setting, parameter inference must be extended to estimate coupling parameters and the underlying network structure. Developing computationally efficient filtering-based methods for this higher-dimensional inference problem is an important direction for future work. Such methods could enable the analysis of multichannel EEG recordings and may be particularly relevant for investigating the spatial propagation and network dynamics of epileptic activity.

More generally, the methodology developed here is not inherently restricted to the stochastic JR-NMM. Similar splitting-based EKF constructions may apply to other Hamiltonian-type or, more broadly, semi-linear SDEs for which suitable structure-preserving numerical splitting methods can be constructed. Extending the approach to such models would provide an opportunity to investigate more systematically how the numerical treatment of stochastic dynamics influences the accuracy and computational efficiency of likelihood-based parameter inference.

Overall, our results demonstrate that combining structure-preserving numerical splitting methods with EKF-based approximate likelihood inference provides an accurate and computationally efficient approach to parameter estimation for the stochastic JR-NMM. The favorable performance of the OU-EKF in simulations and its successful application to real EEG data motivate further developments of splitting-based filtering methods for more complex neural mass models and other partially observed stochastic dynamical systems.

\vspace{1.0cm}
\noindent\textbf{Funding}

\medskip

\noindent The authors declare that they received no specific funding for this research.

\vspace{0.7cm}
\noindent\textbf{Conflict of interest}

\medskip

\noindent The authors declare no conflicts of interest.

\vspace{0.7cm}
\noindent\textbf{Declaration of Generative AI and AI-assisted technologies in the manuscript preparation process}

\medskip

\noindent During the preparation of this work, the authors used OpenAI ChatGPT to assist with language refinement and Anthropic Claude to assist with coding parts of the figure generation. After using these tools, the authors critically reviewed, verified, and edited the content as needed. The authors developed the mathematical derivations and code underlying the proposed methods independently of these tools. The authors take full responsibility for the content of this publication.

\vspace{0.7cm}
\noindent\textbf{Data availability}

\medskip

\noindent The EEG data used in Section~\ref{sec:application} are publicly available at \url{https://repositori.upf.edu/items/77d77b11-1254-4fc2-9065-e37e6ff6e217}. Specifically, we used EEG recording \texttt{O019} in the present study.

\vspace{0.7cm}
\noindent\textbf{Software}

\medskip

\noindent We implemented the EKF-based parameter estimation methods presented in this work in \texttt{R} \citep{Rlang}. Most algorithmic components are written in \texttt{Rcpp} \citep{RcppPack}, which integrates \texttt{C++} into \texttt{R}. We used \texttt{RcppArmadillo} to simplify matrix handling \citep{armadilloPack}. We generated random numbers for the simulations using the \texttt{mvnfast} package \citep{mvnfastPack}, and created images with \texttt{ggplot2} \citep{ggplotPack}. The numerical experiments were performed using the high-performance computing infrastructure of Johannes Kepler University Linz (JKU). Example code is available on \texttt{GitHub} at \url{https://github.com/PauliSimon/OU-EKF-for-JR-NMM.git}.


\end{document}